\documentclass[pdftex]{pasj01}
\usepackage{color}
\usepackage{multirow}
\usepackage[switch,mathlines]{lineno}

\Received{}%{yyyy/mm/dd}
\Accepted{}%{yyyy/mm/dd}
\begin{document} 

\title{ 
%\LETTERLABEL %%% <- uncomment for LETTER article  
%\REVIEWLABEL %%% <- uncomment for REVIEW article  
Measurement of Temperature Relaxation in the Postshock Plasma of the Northwestern Limb of SN~1006}

%%% begin:list of authors
% Do NOT capitalize all letters in "textsc".
\author{Masahiro \textsc{Ichihashi}\altaffilmark{1}}
\altaffiltext{1}{Department of Physics, Graduate School of Science, The University of Tokyo, 7-3-1 Hongo, Bunkyo-ku, Tokyo 113-0033, Japan}
\email{masahiro.ichihashi@phys.s.u-tokyo.ac.jp}

\author{Aya \textsc{Bamba}\altaffilmark{1,2,3}}
\altaffiltext{2}{Research Center for the Early Universe, School of Science, The University of Tokyo, 7-3-1 Hongo, Bunkyo-ku, Tokyo 113-0033, Japan}
\altaffiltext{3}{Trans-Scale Quantum Science Institute, The University of Tokyo, Tokyo  113-0033, Japan}

\author{Yuichi \textsc{Kato}\altaffilmark{1}}

\author{Satoru \textsc{Katsuda}\altaffilmark{4}}
\altaffiltext{4}{Graduate School of Science and Engineering, Saitama University, Shimo-Okubo 255, Sakura, Saitama 338-8570, Japan}

\author{Hiromasa \textsc{Suzuki}\altaffilmark{5, 6}}
\altaffiltext{5}{Department of Physics, Konan University, 8-9-1 Okamoto, Higashinada-ku, Kobe 658-8501, Japan}
\altaffiltext{6}{Institute of Space and Astronautical Science, JAXA, 3-1-1 Yoshinodai, Sagamihara, Kanagawa 229-8510, Japan}

\author{Tomoaki \textsc{Kasuga}\altaffilmark{1}}

\author{Hirokazu \textsc{Odaka}\altaffilmark{1, 7, 8}}
\altaffiltext{7}{Kavli IPMU (WPI), UTIAS, The University of Tokyo, 5-1-5 Kashiwanoha, Kashiwa, Chiba 277-8583, Japan}
\altaffiltext{8}{Department of Earth and Space Science, Graduate School of Science, Osaka University, 1-1 Machikaneyama-cho, Toyonaka-shi, Osaka 560-0043, Japan}

\author{Kazuhiro \textsc{Nakazawa}\altaffilmark{9}}
\altaffiltext{9}{Kobayashi-Maskawa Institute for the Origin of Particles and the Universe, Nagoya University, Furo-cho Chikusa, Nagoya 464-8602, Japan}
%%% end:list of authors

%% `\KeyWords{}' always has to be placed before ``\maketitle'' 
%%  List of Key Words:  https://academic.oup.com/pasj/pages/Pasj_Keywords 

\maketitle

\begin{abstract}
Heating of charged particles via collisionless shocks, while ubiquitous in the universe, is an intriguing yet puzzling plasma phenomenon. One outstanding question is how electrons and ions approach an equilibrium after they were heated to different immediate-postshock temperatures. In order to fill the significant lack of observational information of the downstream temperature-relaxation process, we observe a thermal-dominant X-ray filament in the northwest of SN~1006 with Chandra.
We divide this region into four layers with a thickness of 15$^{\prime\prime}$ or 0.16 pc each, and fit each spectrum by a non-equilibrium ionization collisional plasma model. 
The electron temperature was found to increase toward downstream from 0.52-0.62 keV to 0.82-0.95 keV on a length scale of 60 arcsec (or 0.64 pc). This electron temperature is lower than thermal relaxation processes via Coulomb scattering, requiring some other effects such as plasma mixture due to turbulence and/or projection effects, etc, which we hope will be resolved with future X-ray calorimeter missions such as XRISM and Athena.
\end{abstract}
%\linenumbers
%\pagewiselinenumbers

\section{Introduction}
There are collisionless shocks with various scales in the universe. The collisionless shocks are  defined as shock waves whose scale length of the transition is much smaller than the collisional mean free path of particles. One of the typical examples is seen in supernova remnants (SNRs). 

As a shock wave passes through, the downstream plasma is heated and compressed. 
In the case of strong shocks in ideal gas, the temperature of each particle species directly behind the collisionless shock wave is $\frac{3}{16}m_iv_s^2$, where $m_i$ is the mass of each particle species and $v_s$ is shock speed. This relationship, called as the Rankine-Hugoniot relation, marginally observed in SN~1987A \citep{2019NatAs...3..236M}. Although \citet{2015A&A...579A..13V} and \citet{2014ApJ...780..136Y} found some deviation from the relation.

In the downstream, the plasma temperature gradually approaches the equilibrium after different particle species reach different temperatures. The simplest relaxation process is the Coulomb scattering \citep{1978ppim.book.....S}, although it is yet to be confirmed by observations.
A sign of equilibration due to the Coulomb scattering was observed in the ejecta of Puppis A in a scale of $1'$ or 0.4 pc \citep{2013ApJ...768..182K}. 
However, the relaxation dynamics was not discussed due to the lack of spatial resolution. We thus need to resolve the thermal conditions to much smaller spatial scales.

 In this study, we focus on the SNR SN~1006, which is one of the youngest Galactic type Ia SNRs. 
 The northwestern limb is bright in thermal X-rays \citep{1995Natur.378..255K,2008PASJ...60S.153B,2013ApJ...764..156W,2013ApJ...771...56U,2016MNRAS.462..158L}. With a rather small distance of 2.18\;kpc \citep{2003ApJ...585..324W} or $\sim$1.5\;kpc \citep{2017hsn..book...63K}, Chandra can resolve the filament structure down to $\sim$0.005 pc. This makes this remnant ideal for studying the temperature structure behind the shocks in detail. 
 There are some previous studies of this region. The first observation of emission lines from heavy elements in the ultraviolet spectrum is reported by \citet{1995ApJ...454L..31R}, who suggested that plasma turbulence does not equilibrate electron and proton temperatures within the shock front. The X-ray emission hardens downward \citep{2013ApJ...763...85K} .
 This structure can be at least qualitatively explained by the Coulomb scattering. In order to judge whether the electron heating is simply by the Coulomb scattering, it should be measured the electron temperature as a function of the distance from the shock front.

This paper consists of the following sections: section~2 describes the observation and data reduction. Section~3 summarizes the result of the spatially resolved spectral analysis. In section~4, we discuss the origin of the spatial variation of the electron temperature and note the possibility of the metal abundance variation. The errors in this paper are set to 90\% confidence level.

\section{Observation and Data Reduction}
Chandra has observed the northwestern region of SN~1006 several times. We use two data sets with longest exposures; ObsID 1959 (observed in 2001) and ObsID 13737 (observed in 2012) (see Table \ref{tab:data}). 
These data are reprocessed and analyzed with the software packages Chandra Interactive Analysis of Observation (CIAO) version 4.13.0 \citep{2006SPIE.6270E..1VF}, CALDB 4.9.4 and XSPEC software version 12.11.1 \citep{1996ASPC..101...17A}. The resultant exposure is 88.98\;ks (ObsID 1959) and 87.09\;ks (ObsID 13737). We use the C-statistic \citep{1979ApJ...228..939C} for our analysis below.

\begin{table}
\tbl{List of obtained data sets}{%
\begin{tabular}{ccc}
\hline
ObsID&Obs. Start Date&Exposure\\
&(YYYY-MM-DD)&(ks)\\
\hline\hline
1959&2001-04-26&88.98\\
13737&2012-04-20&87.09\\
\hline
\end{tabular}}\label{tab:data}
\begin{tabnote}
\end{tabnote}
\end{table}

\begin{figure*}
\begin{tabular}{cc}
\begin{minipage}[c]{0.5\linewidth}
\includegraphics[keepaspectratio,width=1.0\columnwidth]{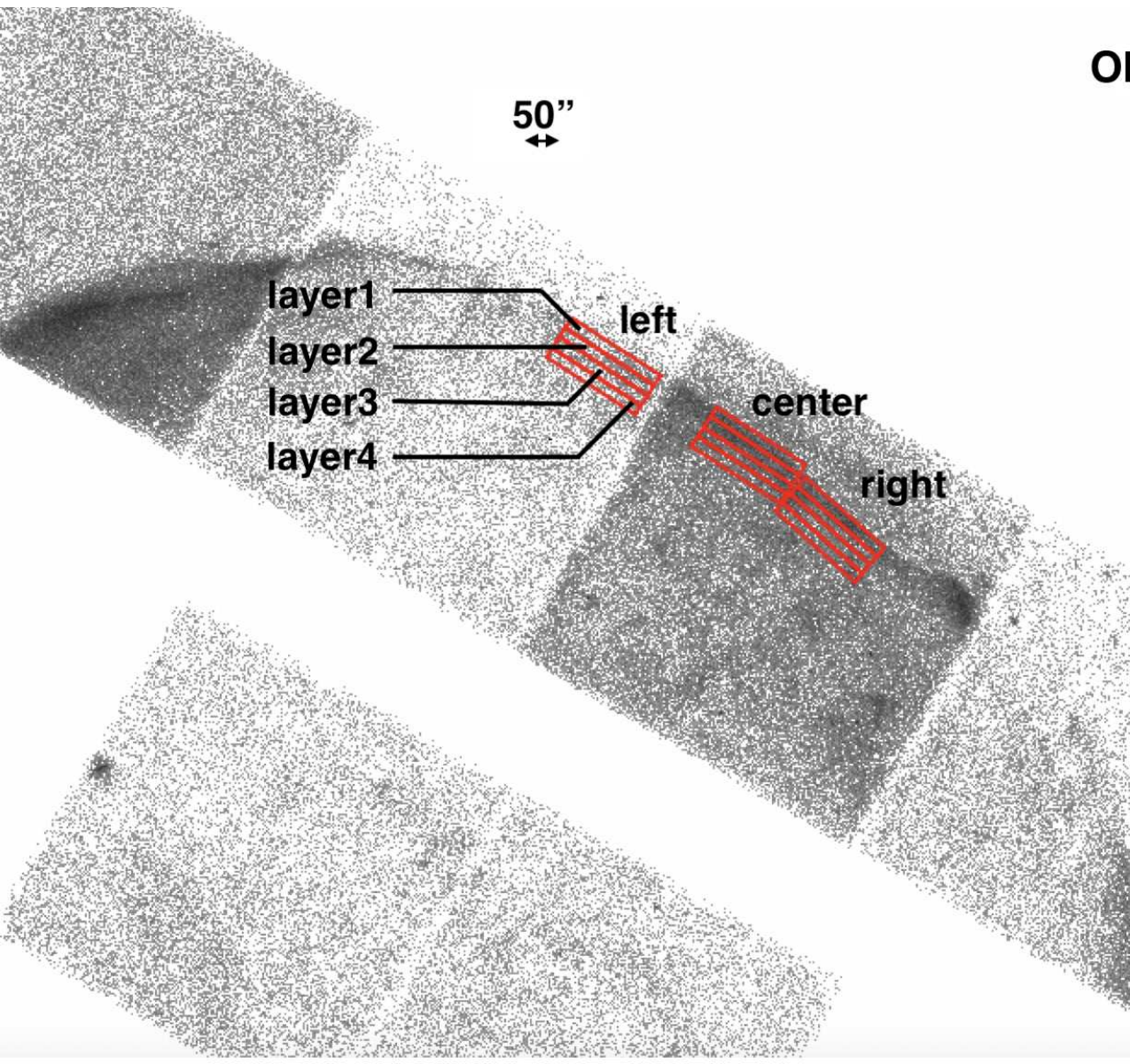}
\end{minipage}
\begin{minipage}[c]{0.5\linewidth}
\includegraphics[keepaspectratio,width=1.0\columnwidth]{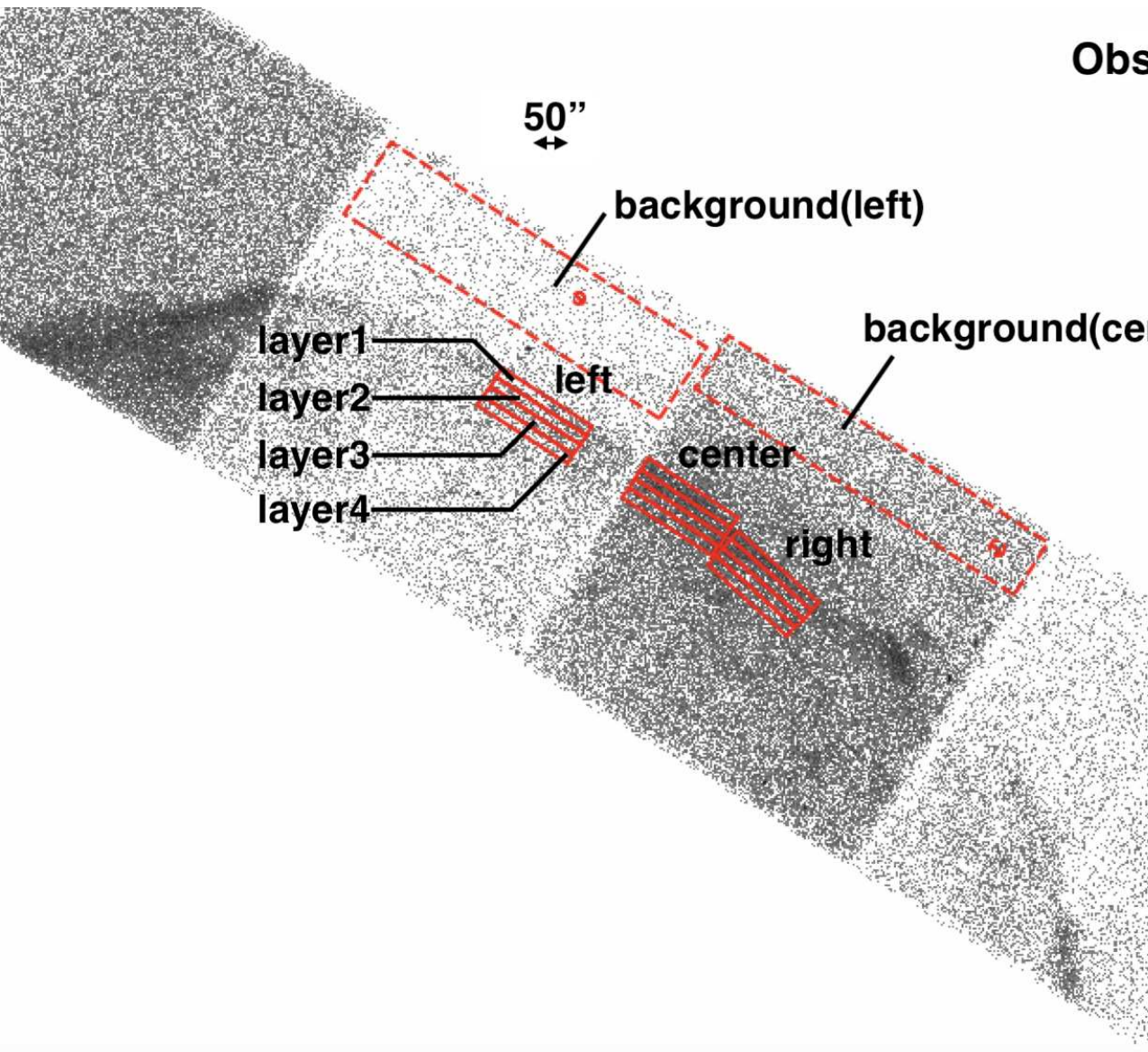}
\end{minipage}
\end{tabular}
\hspace{-10truemm}
\caption{The 0.5-2.0~keV images of the northwestern region of SN~1006 observed by \textit{Chandra} (ObsID : 1959 (left), 13737 (right)). Both images are shown on a log scale. Solid and dashed regions represent the source and background regions for the spectral analysis (see text). For each of the left, center, and right regions, the four layers from the filament toward downstream are labeled as layer 1, 2, 3, and 4 as described for the left regions. Two bright point sources shown are excluded from the background region of the data observed in 2012. In analysis, we make the background model based on the background of 13737 (see \S 3.1).}\label{fig:region}
\end{figure*}

\section{Analysis and results}
\subsection{Background estimation}
First, we obtain background spectrum from the source-free region as shown in Figure \ref{fig:region}. Since our source regions are covered by both BI (back-illuminated) and FI (front-illuminated) CCDs, we have checked the background spectra in both types of CCDs separately. Two bright point sources shown in Figure \ref{fig:region} are excluded from the background region of the data of ObsID 13737. The observation of ObsID 1959 was fully occupied by the emission from SN~1006, and thus we cannot select background regions. The simplest way of estimating background of ObsID 1959 is simply substituting the background of ObsID 13737, but such a simple replacement is inappropriate because of the several reasons. We therefore developed a model background, as described below.

The X-ray background can be divided into two main components, the particle induced background (PIB) and the sky background (\cite{2021A&A...655A.116S}). PIB is the background induced by cosmic-ray particles, which depends on  both the solar activity and the property of the detector. In this analysis, PIB was modeled by using {\tt mkacispback} package (\cite{2021A&A...655A.116S}). We make a PIB model for eight regions; center, right, and left regions of both  observation data and two background regions of the data of ObsID 13737 (see Figure \ref{fig:region}).

The sky background is the emission from celestial sources. 
First, we consider four potential origins of the sky background: the Milky Way Halo (MWH), the local excess (possibly nearby supernova explosion \citep{1994PASJ...46..367O}), and the Lupus Loop (hot and cold components), following \citet{2013ApJ...771...56U}. The MWH and the Lupus Loop have old plasma and are thought to reach ionization equilibrium. In this model, we thus consider MWH by the XSPEC model {\tt equil} with fixed electron temperature of 0.1 keV, local excess by the XSPEC model {\tt powerlaw}, and the hot and cold components of the Lupus Loop by two XSPEC models {\tt equil}. As for the interstellar absorption, we use the XSPEC model {\tt phabs} with the hydrogen column densities fixed to $5.6 \times 10^{20}\;\mathrm{cm^{-2}}$ for MWH and $1.0 \times 10^{20}\;\mathrm{cm^{-2}}$ for the others, which are the same as those used in \citet{2013ApJ...771...56U}. As a result, the normalization of the MWH components is found to be consistent with zero. Also, the temperatures of two Lupus Loop components are consistent with each other, and thus one {\tt equil} model is sufficient. This result implies that the sky background in this region can be well reproduced with two components: the local excess and the Lupus Loop component. 

\begin{figure*}
\begin{minipage}[b]{0.5\linewidth}
\includegraphics[keepaspectratio,width=\columnwidth]{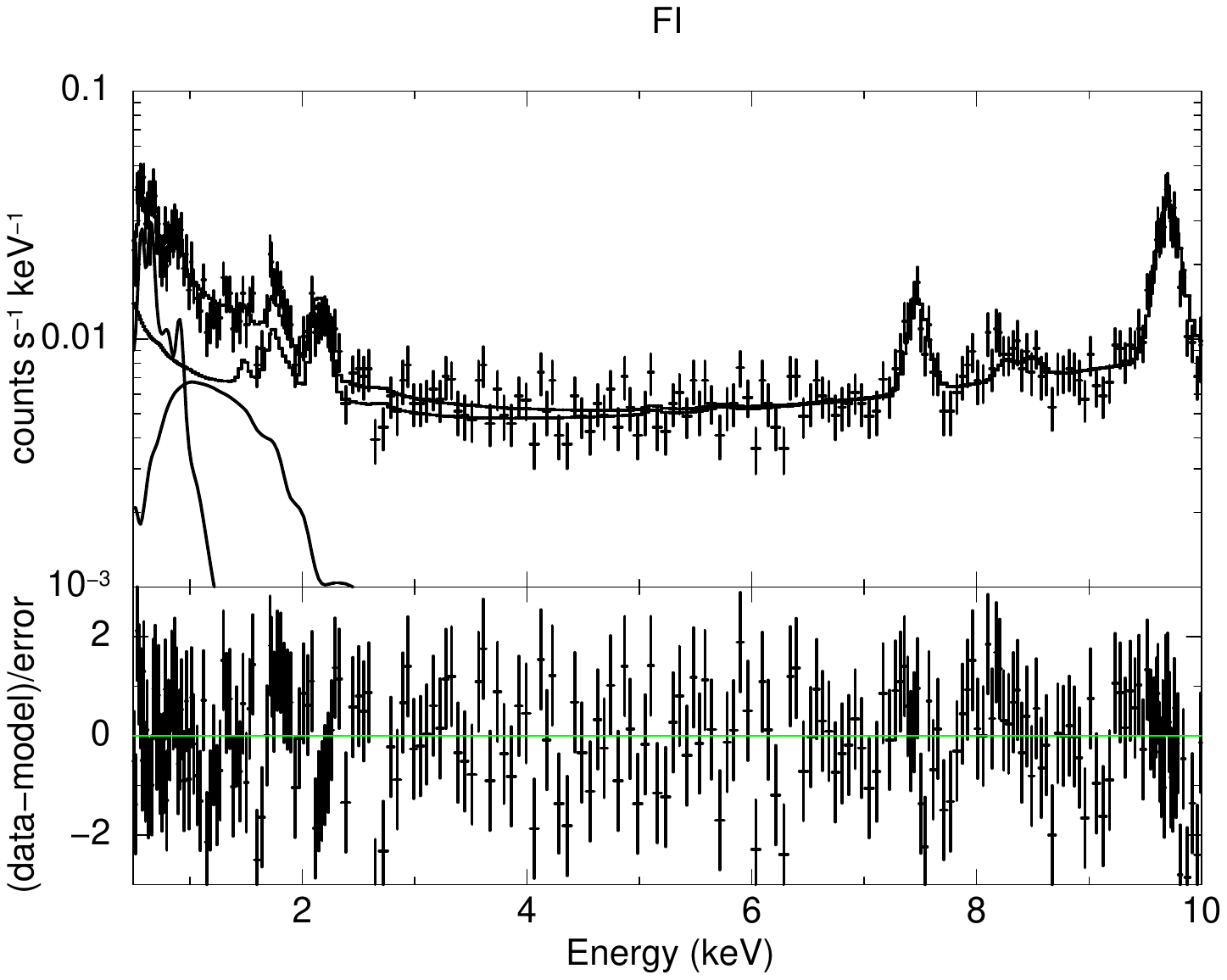}
\end{minipage}
\begin{minipage}[b]{0.5\linewidth}
\includegraphics[keepaspectratio,width=\columnwidth]{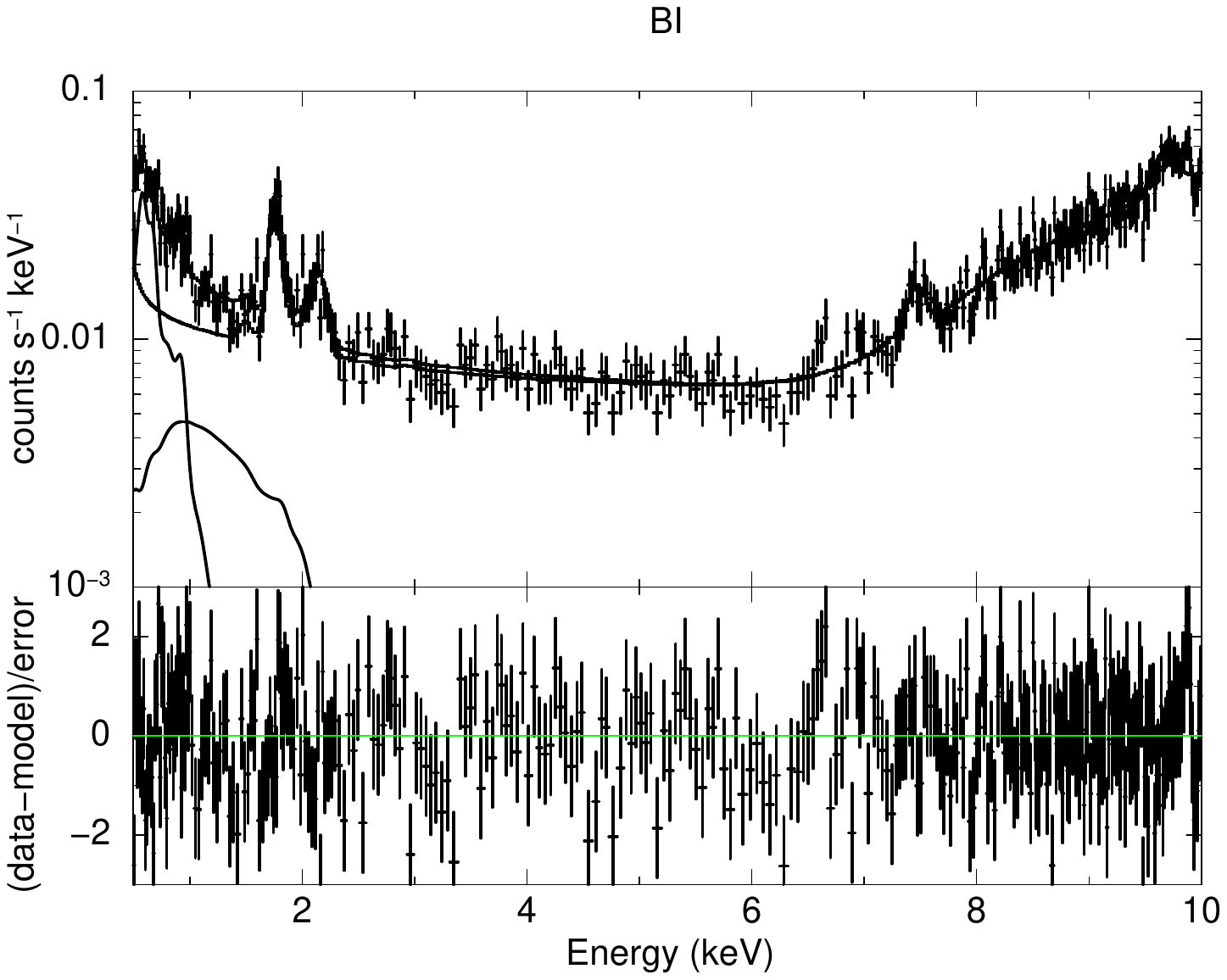}
\end{minipage}
\caption{Background spectra on ObsID 13737 and the best-fit models. Left and right panels show the spectra extracted from the FI and BI CCDs, respectively. The dashed, dotted, and dash-dotted lines represent the particle induced background (PIB), the local excess and the Lupus Loop, respectively.}\label{fig:bestfit_sky}
\end{figure*}

We fit the sky background spectra with a model composed of these components. The best-fit parameters and models are shown in Table \ref{tab:bestfit_sky} and Figure \ref{fig:bestfit_sky}. The sky background is assumed to be uniform over our background and source regions and its parameters are fixed to the values obtained above (Table \ref{tab:bestfit_sky}) for the study of the source regions described below.

\begin{longtable}{*{4}{c}}
\caption{Best-fit Parameters of sky background}
\label{tab:bestfit_sky}
\hline
\multicolumn{2}{c}{Parameter}&\multicolumn{2}{c}{Value} \\
&&BI CCD&FI CCD\\
\endhead
\hline
Absorption {\tt (phabs)}& $N_{\mathrm{H}}\;(\times10^{20}\mathrm{cm^{-2}})$ &\multicolumn{2}{c}{1.0 (fix)}  \\
\hline
local excess {\tt(powerlaw)}&photon index  &\multicolumn{2}{c}{$2.21_{-0.41}^{+0.52}$}\\
&norm (photons/keV/cm$^2$/s)$^a$&$1.42_{-0.39}^{+0.47} \times10^{-5}$&$2.77_{-0.53}^{+0.63}\times10^{-5}$\\
Lupus Loop {\tt(equil)}&$kT\;(\mathrm{keV})$&\multicolumn{2}{c}{$0.18_{-0.01}^{+0.01}$}\\
&abundance (/solar)&\multicolumn{2}{c}{0.20 (fix)}\\
&norm  ($\mathrm{cm}^{-2}$)&$5.20_{-0.58}^{+0.65}\times10^{-4}$&$9.12_{-1.24}^{+1.41}\times10^{-4}$\\
\hline\hline
\multicolumn{2}{c}{c-stat/d.o.f}&\multicolumn{2}{c}{$1344.68/1294$}\\
\hline
\multicolumn{4}{l}{$a$: at 1 keV}\\

\end{longtable}

\subsection{Spectral analysis on the source regions}
We divide the northwestern region of SN~1006 into three regions, which are labeled as center, right, and left (see Figure \ref{fig:region}). Each of them has a rectangular shape of $140^{\prime\prime}\times60^{\prime\prime}$ or $1.5\; \mathrm{pc} \times0.64\;\mathrm{pc}$ at 2.18\;kpc distance, and its long side coincides with the shock front of each observation. Each region is further divided into four layers with a thickness of $15^{\prime\prime}$ or 0.16 pc each. They are labeled as layer 1, 2, 3 and 4 from the shock front toward downstream (see Figure \ref{fig:region}). We extract spectra from all the 12 regions of the data of both ObsID 1959 and 13737.

\begin{figure*}[htbp]
\begin{tabular}[htbp]{ccc}
\begin{minipage}[t]{.33\textwidth}
\includegraphics[keepaspectratio,width=\columnwidth]{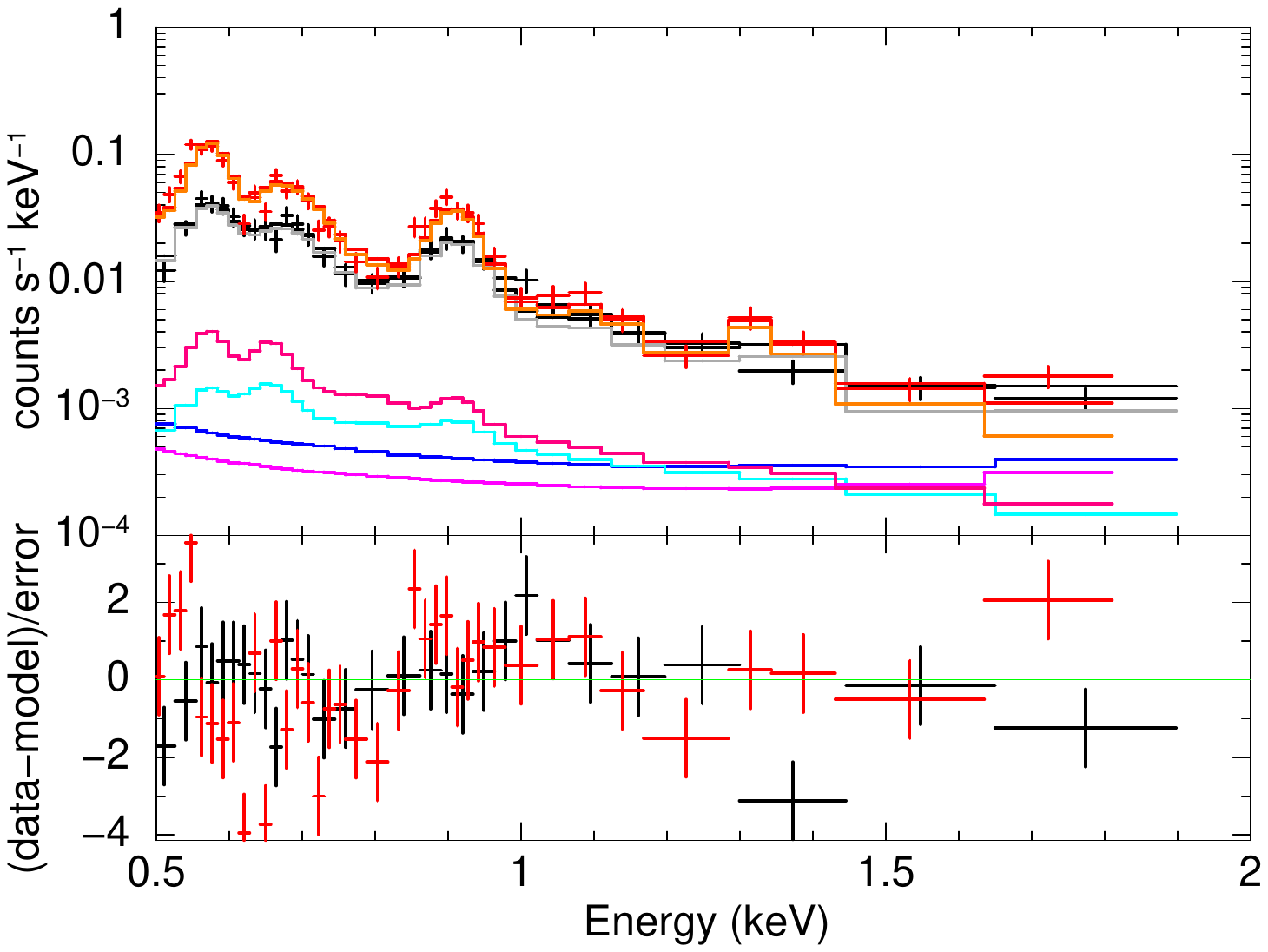}
\end{minipage}&
\begin{minipage}[t]{.33\textwidth}
\includegraphics[keepaspectratio,width=\columnwidth]{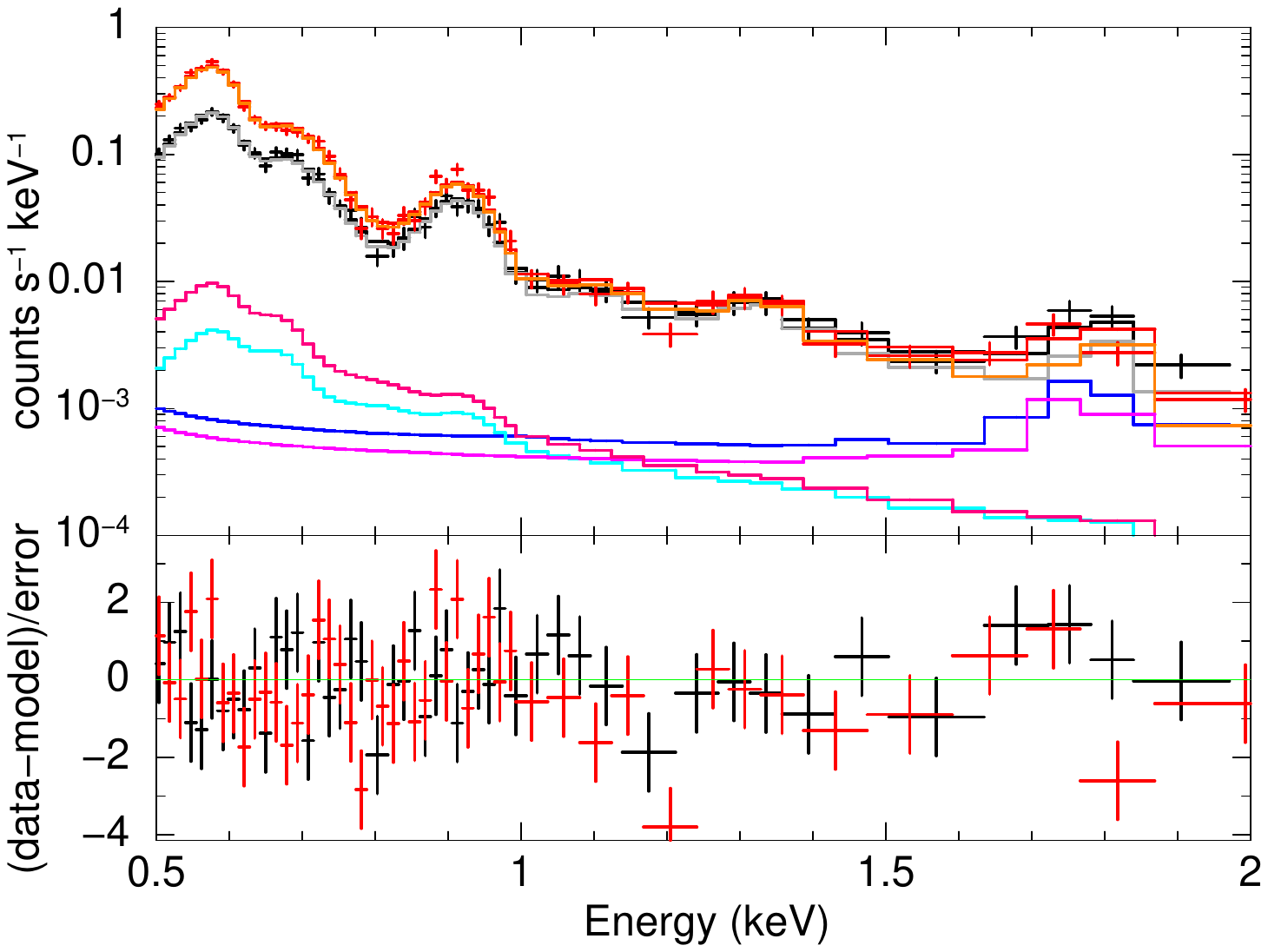}
\end{minipage}&
\begin{minipage}[t]{.33\textwidth}
\includegraphics[keepaspectratio,width=\columnwidth]{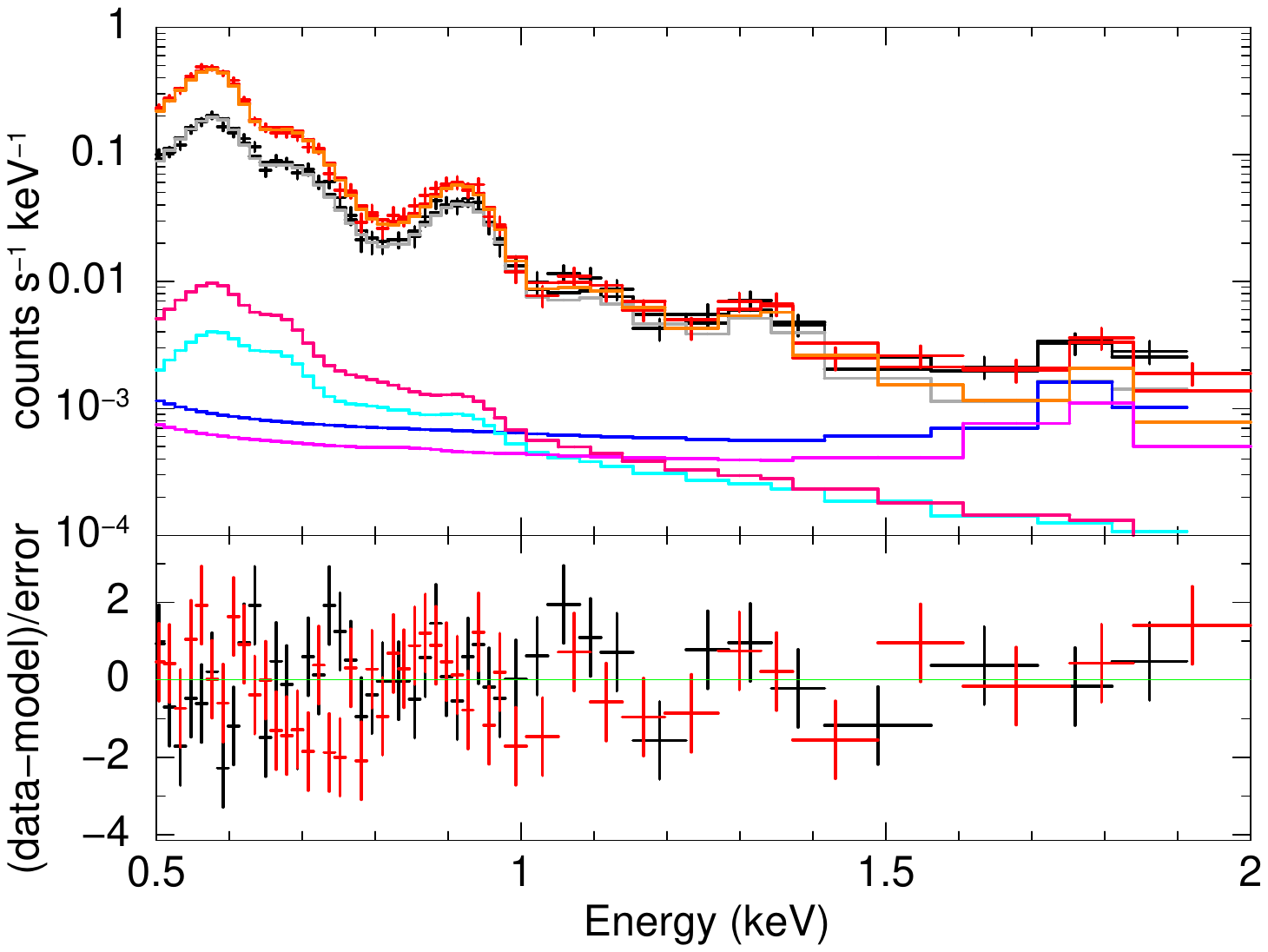}
\end{minipage}\\
\begin{minipage}[t]{.33\textwidth}
\includegraphics[keepaspectratio,width=\columnwidth]{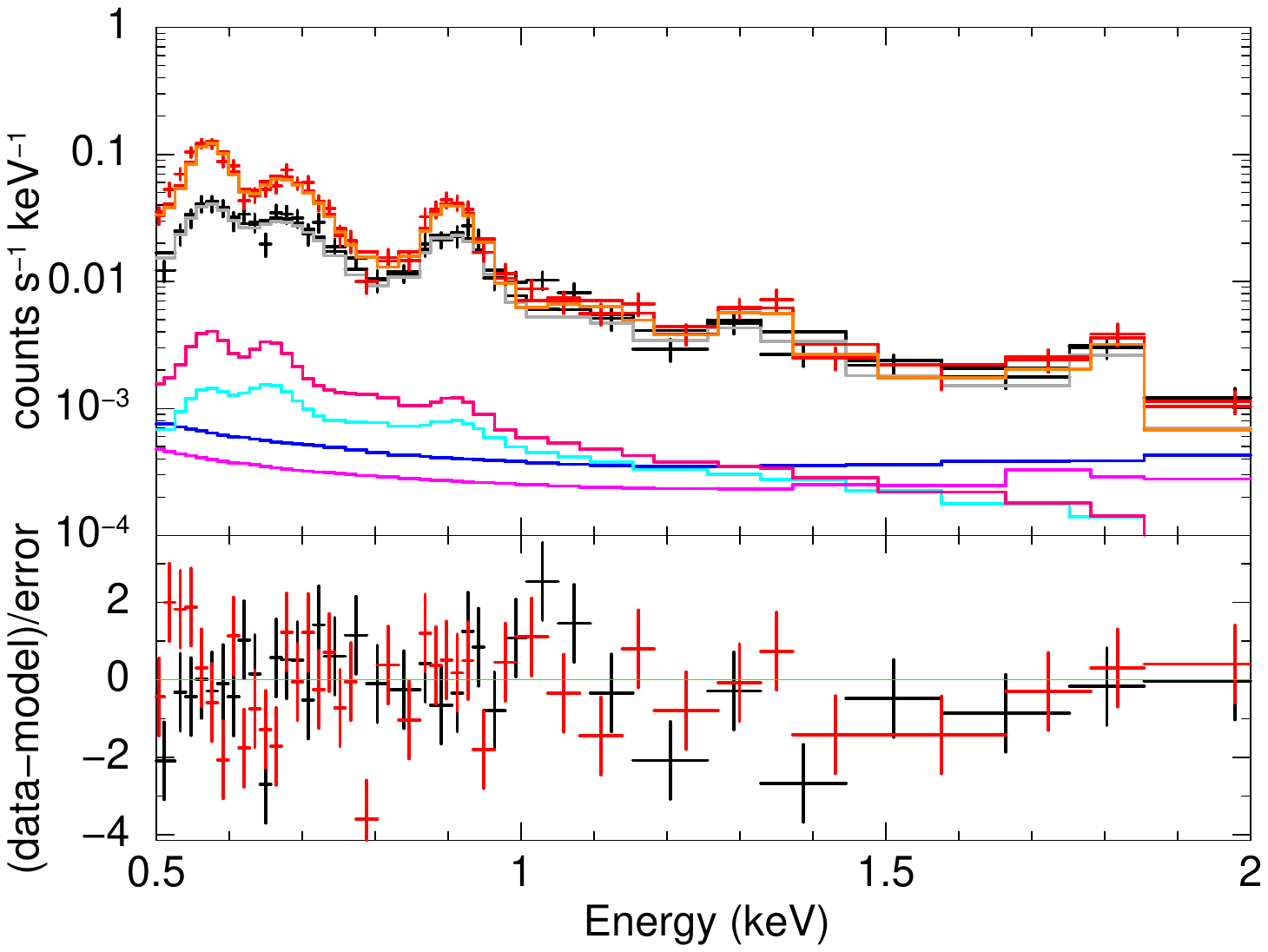}
\end{minipage}&
\begin{minipage}[t]{.33\textwidth}
\includegraphics[keepaspectratio,width=\columnwidth]{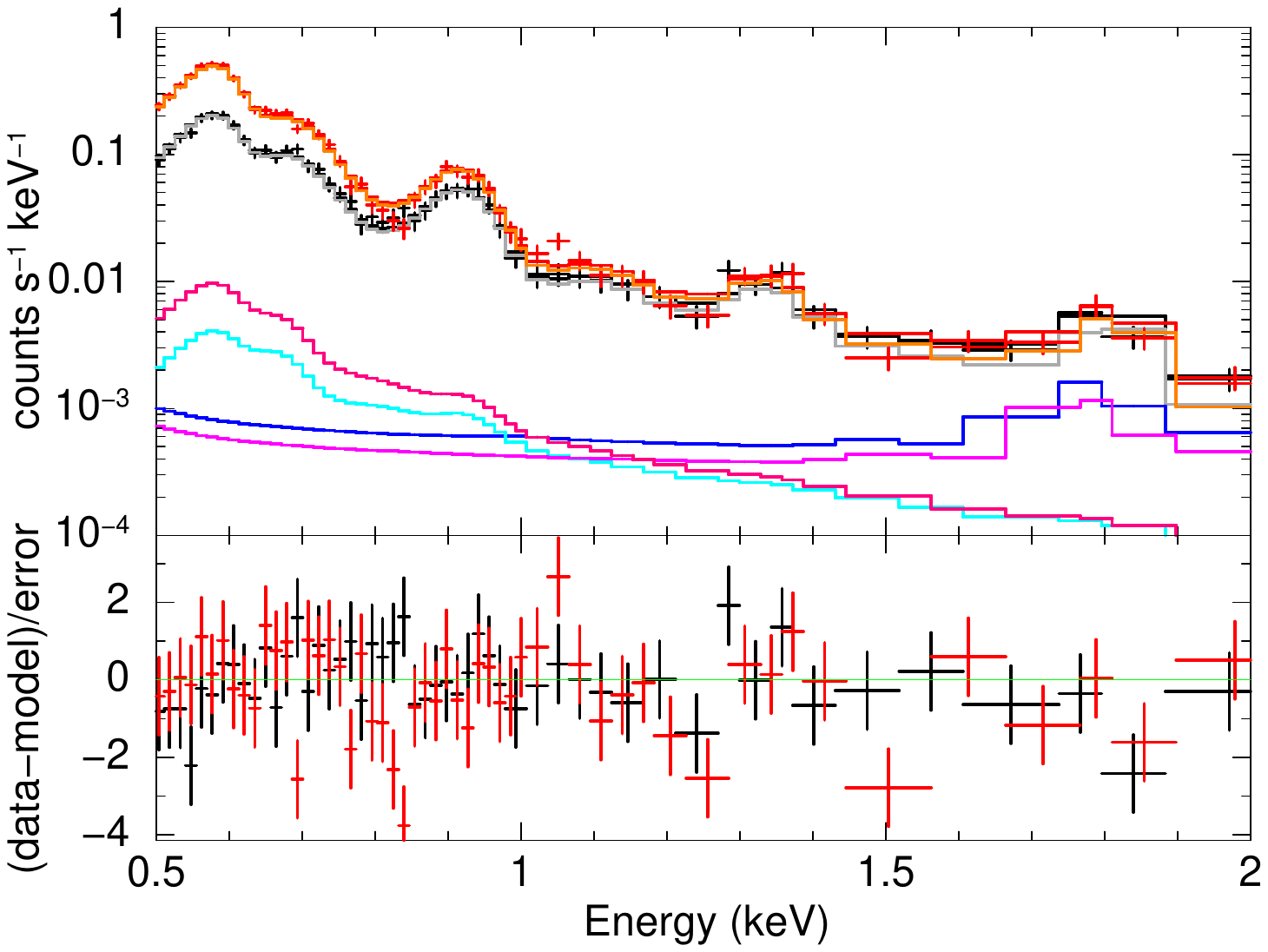}
\end{minipage}&
\begin{minipage}[t]{.33\textwidth}
\includegraphics[keepaspectratio,width=\columnwidth]{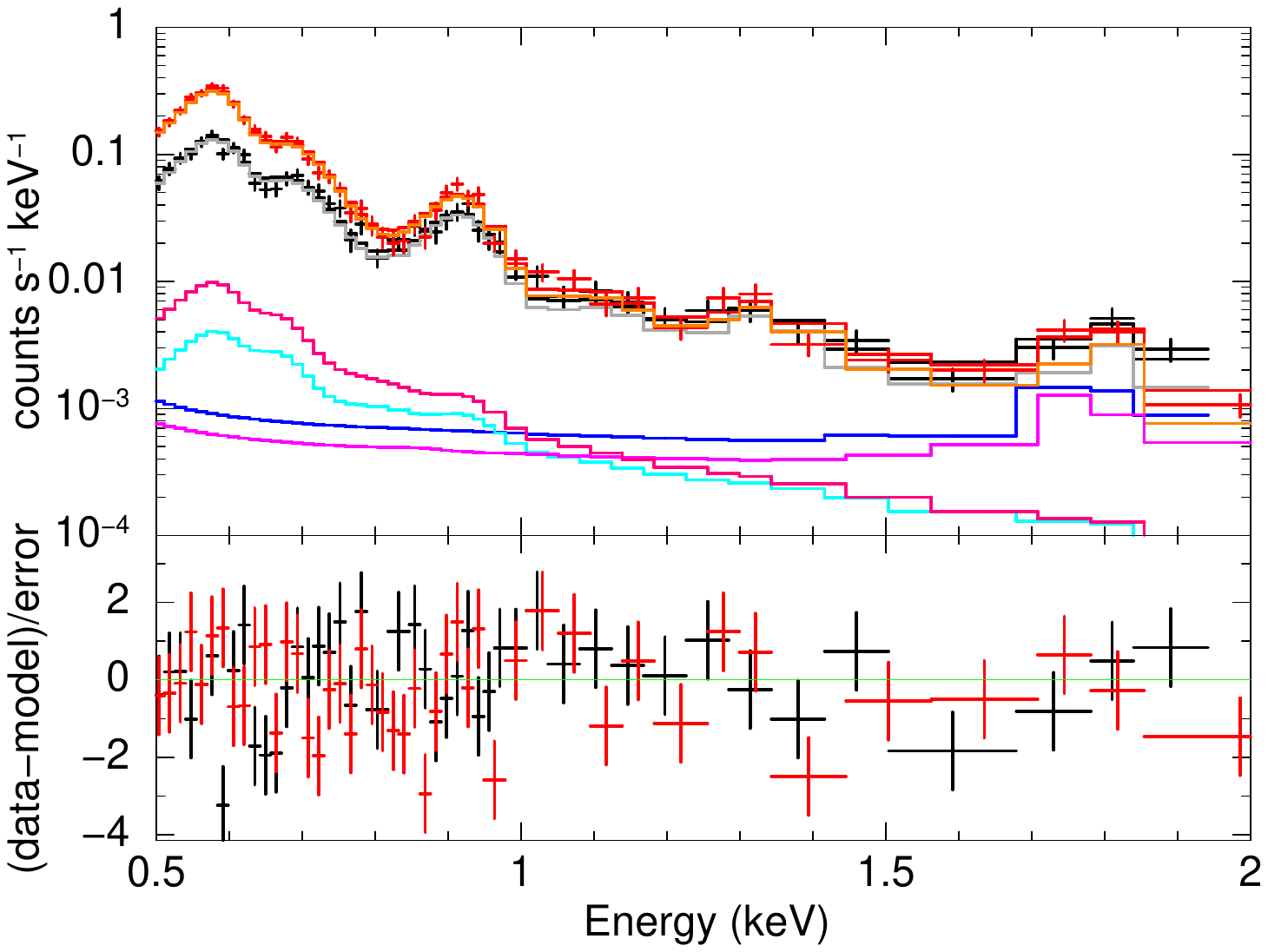}
\end{minipage}\\
\begin{minipage}[t]{.33\textwidth}
\includegraphics[keepaspectratio,width=\columnwidth]{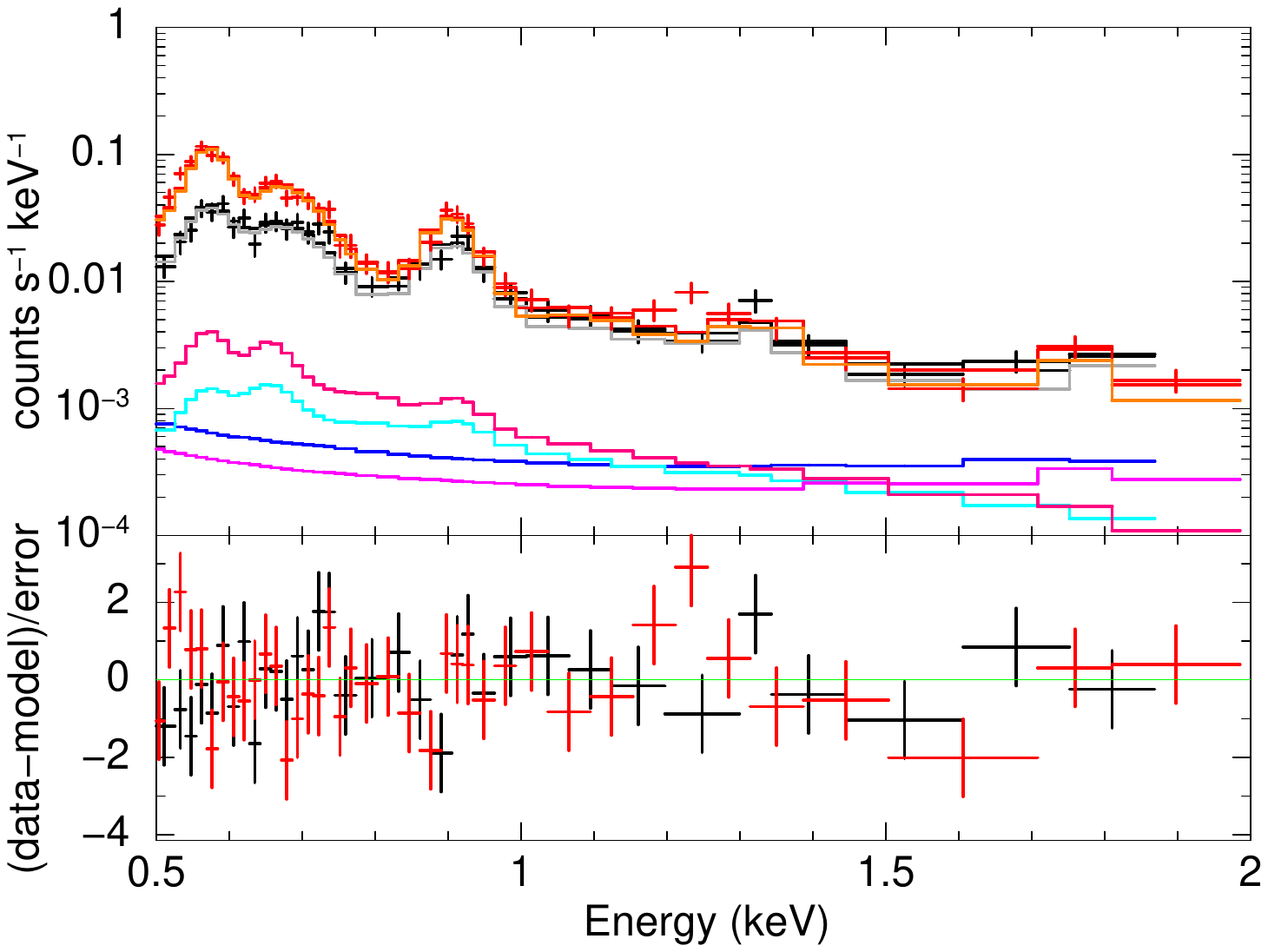}
\end{minipage}&
\begin{minipage}[t]{.33\textwidth}
\includegraphics[keepaspectratio,width=\columnwidth]{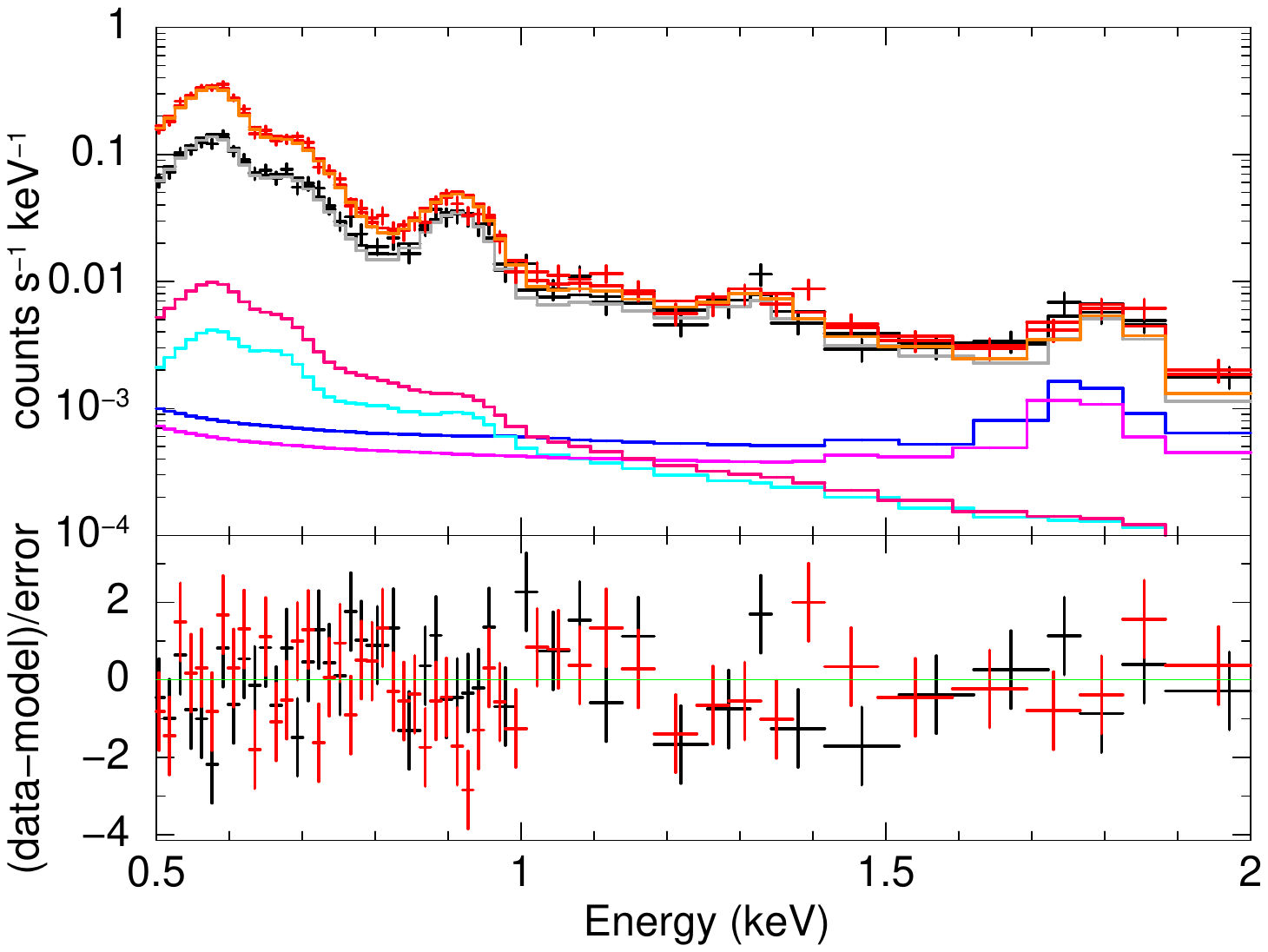}
\end{minipage}&
\begin{minipage}[t]{.33\textwidth}
\includegraphics[keepaspectratio,width=\columnwidth]{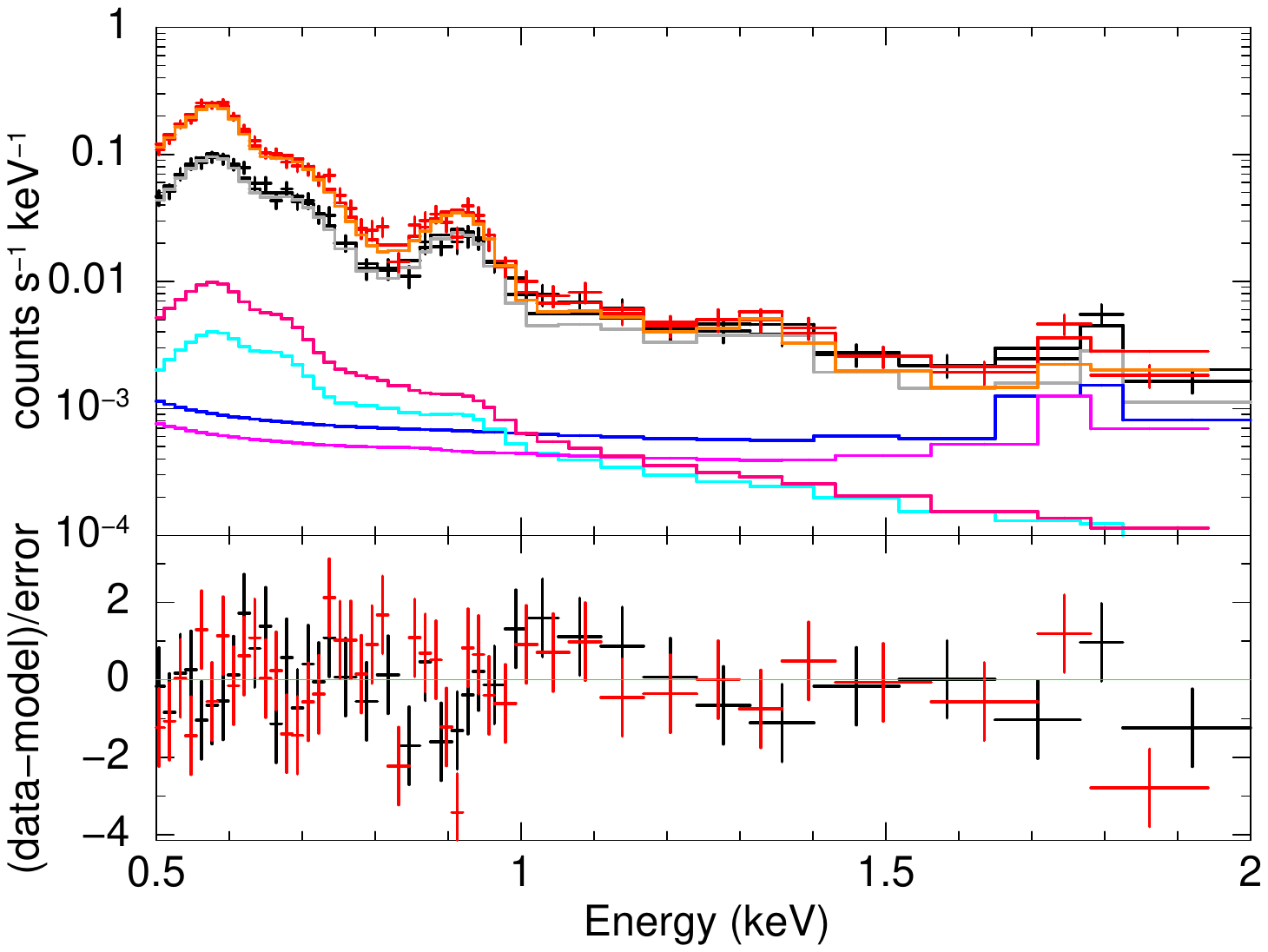}
\end{minipage}\\
\begin{minipage}[t]{.33\textwidth}
\includegraphics[keepaspectratio,width=\columnwidth]{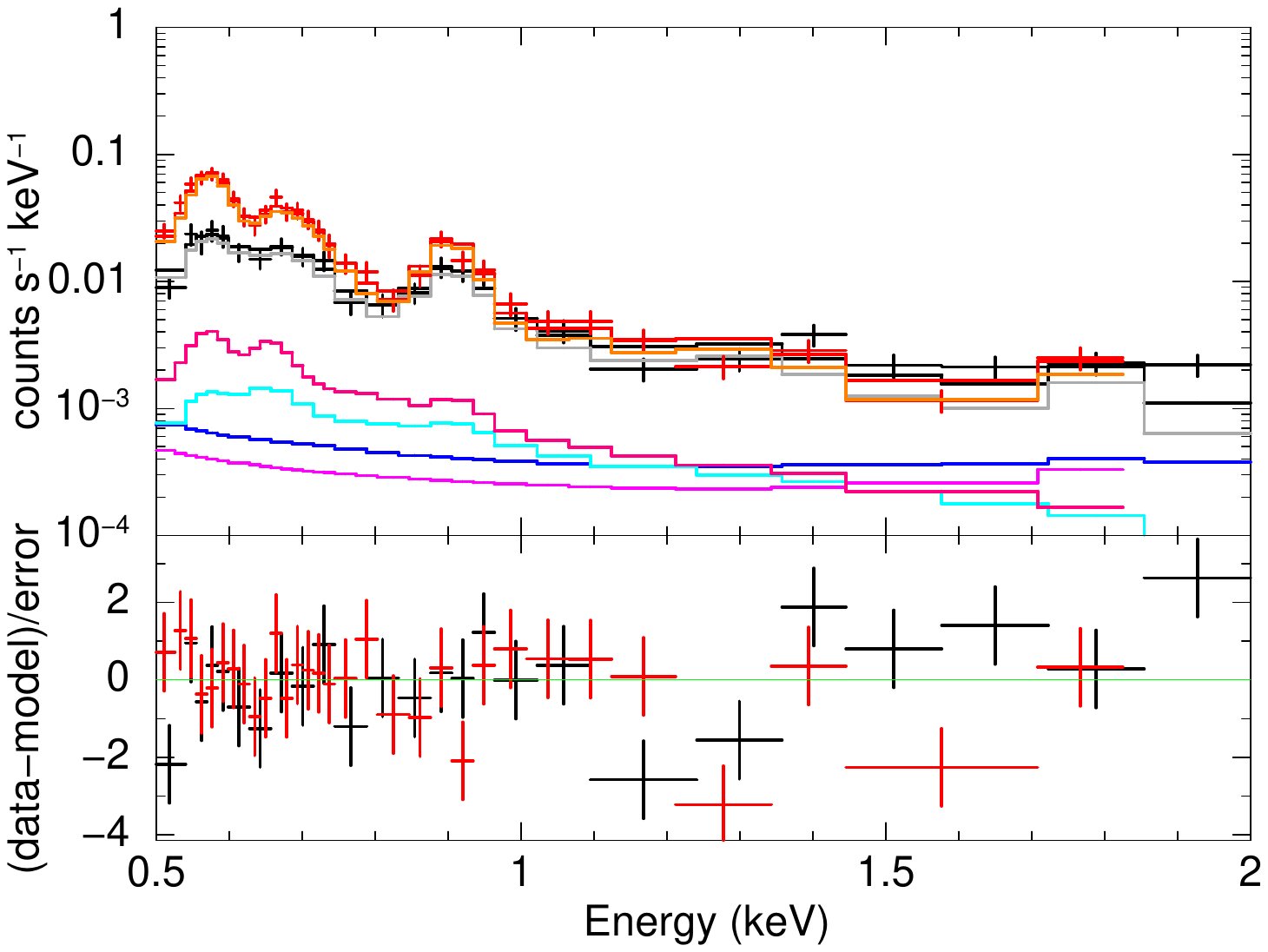}
\end{minipage}&
\begin{minipage}[t]{.33\textwidth}
\includegraphics[keepaspectratio,width=\columnwidth]{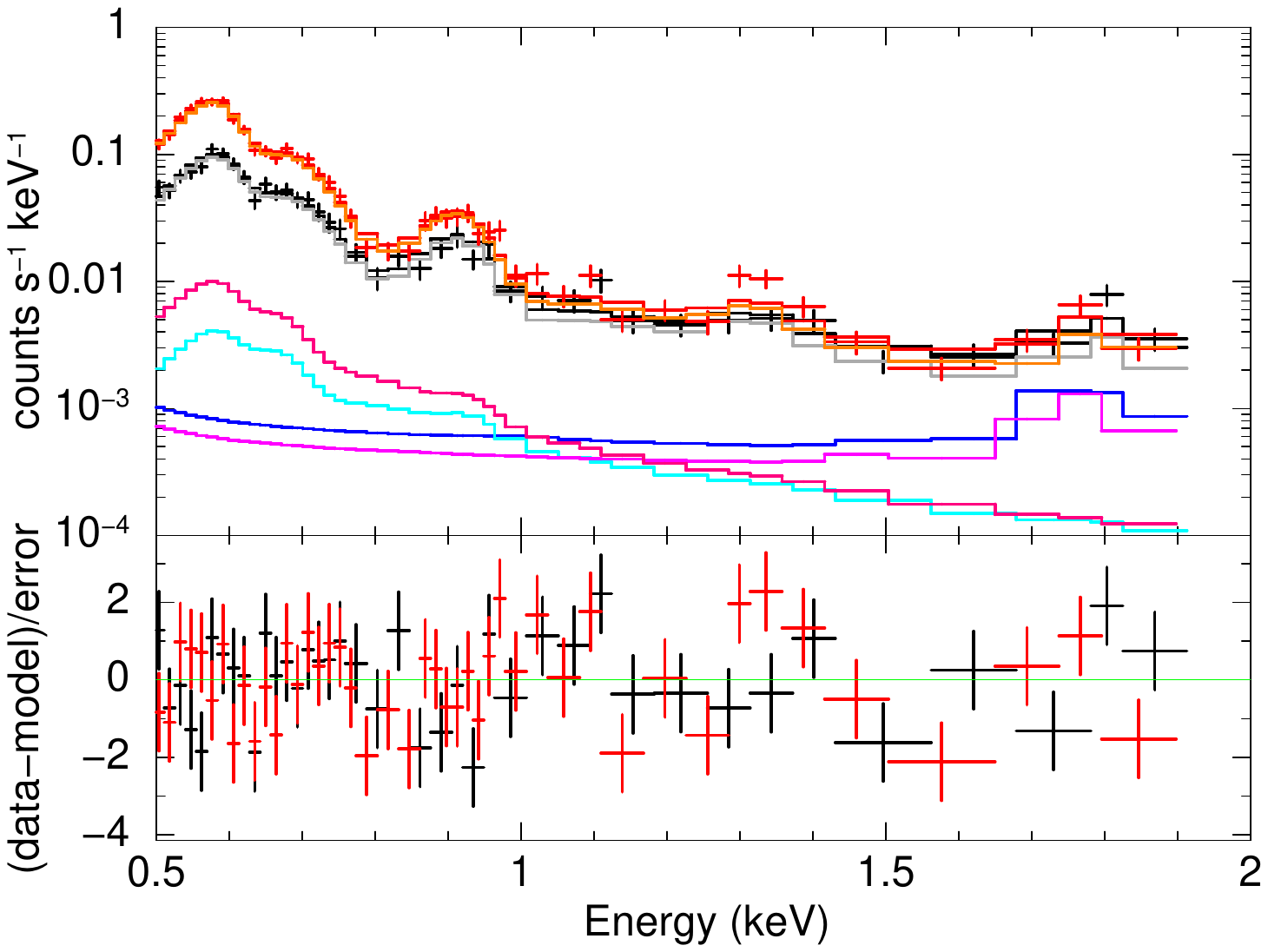}
\end{minipage}&
\begin{minipage}[t]{.33\textwidth}
\includegraphics[keepaspectratio,width=\columnwidth]{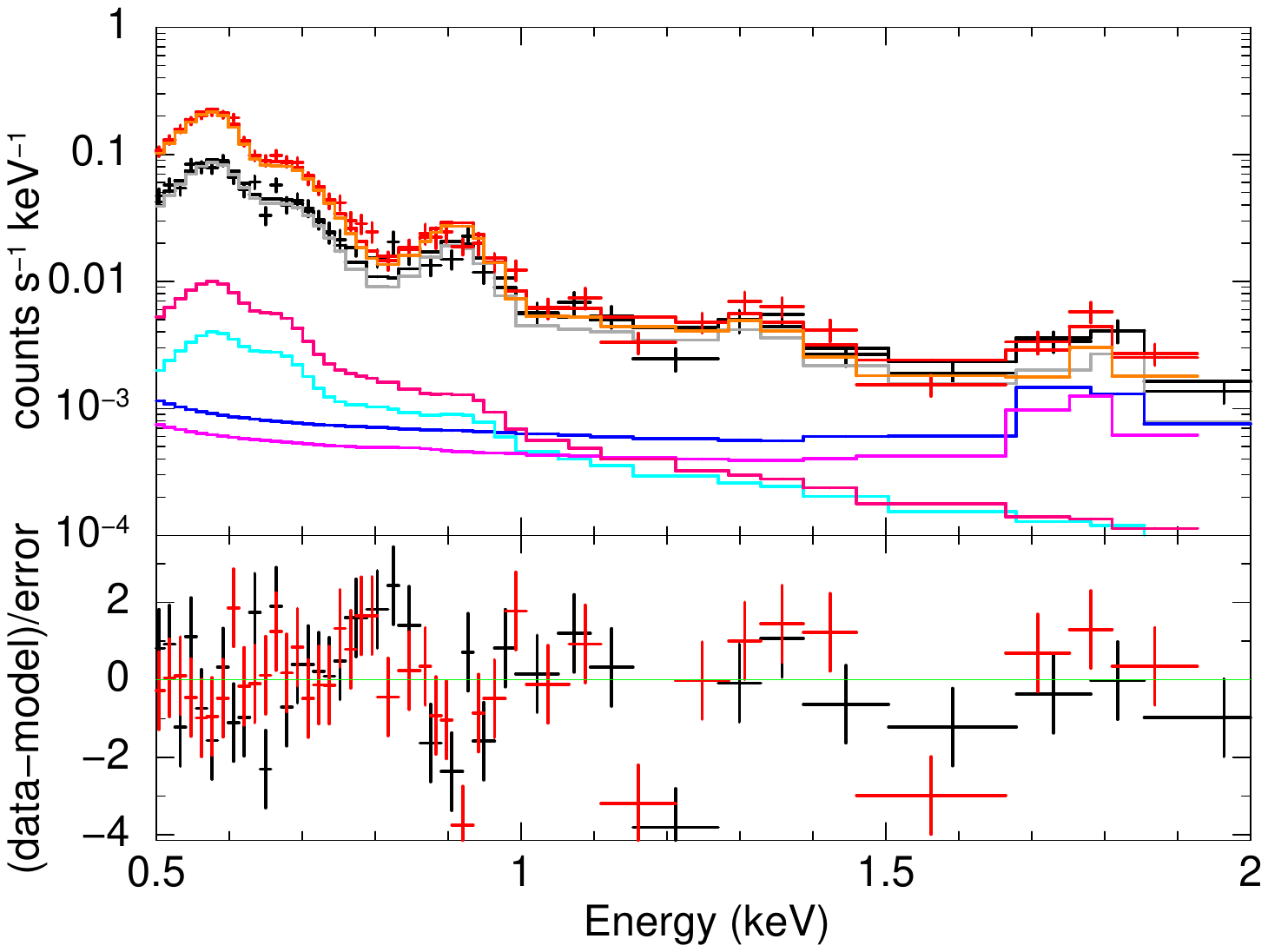}
\end{minipage}\\
\end{tabular}
\caption{Spectra of each analysis region and the best-fit model. Columns mean left, center and right from left to right, and rows mean layer 1,2,3,4 from top to bottom. Red and black crosses show the data of ObsID 1959 and 13737, respectively. Six solid lines show the model components: VNEI of ObsID 1959 (pink) and 13737 (gray), PIB of ObsID 1959 (magenta) and 13737 (blue), and sky background of 1959 (orange) and 13737 (light blue).}\label{fig:fit_vnei}
\end{figure*}

\begin{figure}
\centering
\includegraphics[keepaspectratio,width=\columnwidth]{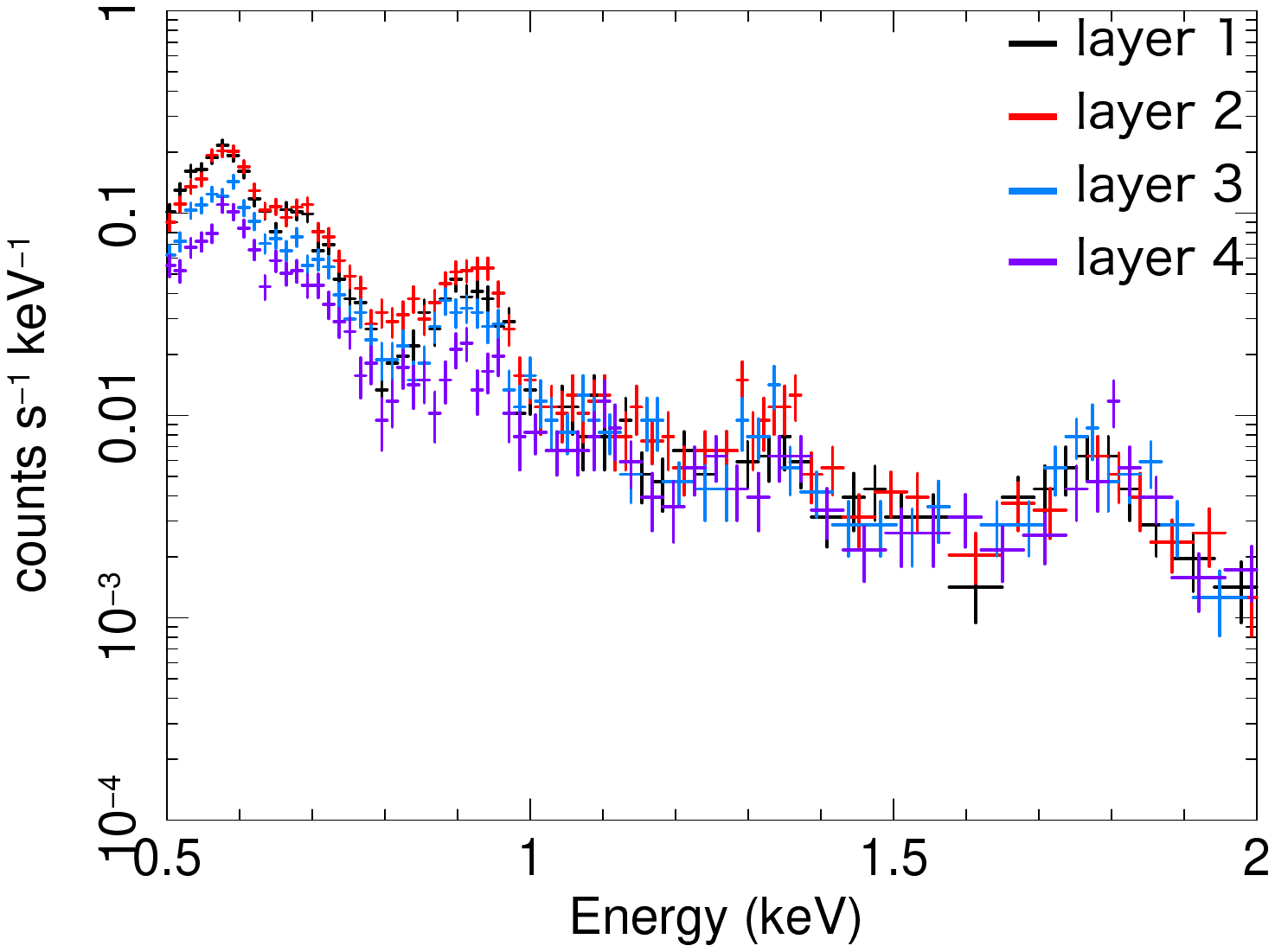}
\caption{Comparison of spectra of center region of ObsID 13737.}\label{fig:compare_spectrum}
\end{figure}

Figure \ref{fig:fit_vnei} shows the spectra from all 12 layers, and Figure \ref{fig:compare_spectrum} shows the comparison of the spectrum from each layer in the center region. 
One can see from Figure \ref{fig:compare_spectrum} that the spectra from layer 1 show softer spectra compared with those in inner layers. This implies that the temperature is higher in the inner regions than the outermost regions. As the spectral model for these source regions, we use a simple model, an absorbed non-equilibrium ionization collisional plasma model with variable abundance (VNEI). We use the abundance derived from \citet{1989GeCoA..53..197A}. We assume that the emission is from the shocked interstellar medium heated by the shockwave, since the region we observed is very close to the shock front and will not be contaminated severely by the ejecta \citep{2013ApJ...771...56U}. Also, we do not consider the non-thermal components from electrons accelerated on the shock front (\cite{1995Natur.378..255K}, \cite{2003ApJ...589..827B}), because the X-ray emission from this part of SN~1006 is mostly from the shock-heated plasma (e.g., \cite{2016MNRAS.462..158L}). 
The abundances of N,  O, Ne, Mg, Si and Fe are free but are tied among all of the layers. All the parameters except for the normalization of VNEI are linked between the two observations, because the conditions of each layer between two observations are considered the same by the correction due to moving each layer along the proper motion.  
We consider the possible change in the brightness of the source emission in time by leaving the normalization of VNEI untied. 
The interstellar absorption is considered by the {\tt phabs} model, for which the hydrogen column density is fixed to $N_{\rm H}=4.16\times10^{20}\;\mathrm{cm^{-2}}$ (\cite{2013A&A...552A...9B}).

\begin{figure}
\centering
\includegraphics[keepaspectratio,width=0.9\columnwidth]{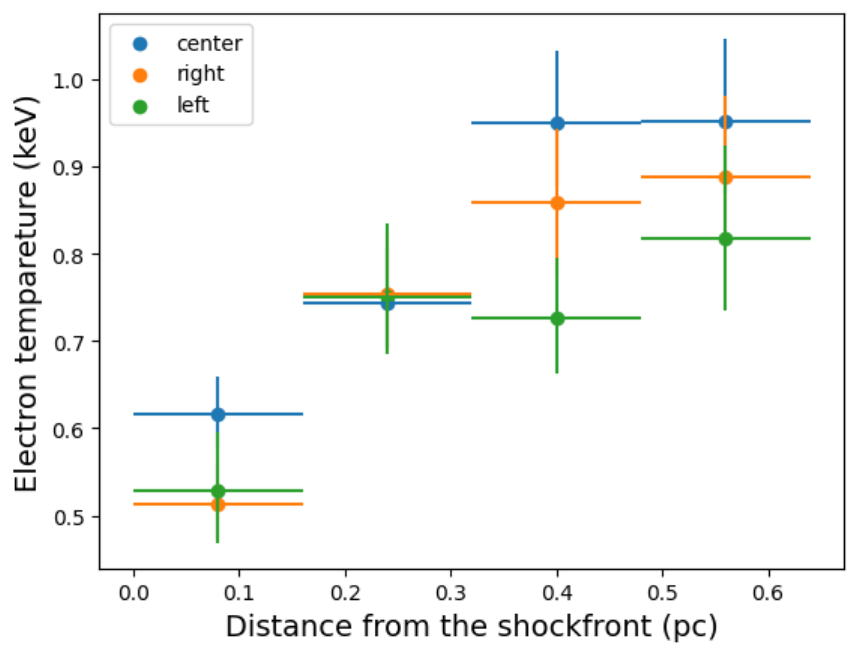}
\caption{Electron temperature $kT_e$ as a function of the distance from the shock front. }\label{fig:kT}
\end{figure}

We fit all of the 12 spectra simultaneously. The fitting result is acceptable with the c-statistics/d.o.f. (degree of freedom) of $16012.98/15564$. The best-fit results are shown in Table \ref{tab:bestfit_vnei} and Figure \ref{fig:fit_vnei}. Note that the sky background level has some uncertainties. We perform the same analysis with the background normalization increased or decreased within their error range compared to the best-fit value, and find no significant difference in our results.
Figure \ref{fig:kT} shows the spatial variation of the electron temperature in each of the left, center and right region. One can see that the temperature gradually increases from the layer 1 to 4.

\begin{longtable}[htbp]{*{5}{c}}
\caption{Best-fit Parameters of spectrum\footnotemark[$*$]}\label{tab:bestfit_vnei}
\hline
\multicolumn{2}{c}{Parameter}&\multicolumn{3}{c}{Value} \\
&&left&center&right\\
\hline
\endhead
\endfoot
\hline
\multicolumn{5}{l}{\footnotemark[$*$] Errors are in 90\% range.}\\ 
\multicolumn{5}{l}{\footnotemark[$\dag$] $\frac{10^{-14}}{4\pi D^2}\int n_{\mathrm{e}}n_H dV$ in the unit of cm$^{-5}$, where $D$ is the angular diameter distance to the source (cm),}\\
 \multicolumn{5}{l}{$dV$ is the volume element (cm$^3$), and $n_{\mathrm{e}}$ and $n_H$ are the electron and H densities (cm$^{-3}$), respectively.}\\
\hline
\endlastfoot
\hline
\multicolumn{5}{c}{common}\\
\hline
Absorption& $N_H\;(\times10^{20}\mathrm{cm^{-2}})$ &\multicolumn{3}{c}{4.16\;(fix)}  \\
abundance\;(/solar)&N&\multicolumn{3}{c}{$0.37_{-0.06}^{+0.07}$}\\
&O&\multicolumn{3}{c}{$0.30_{-0.02}^{+0.02}$}\\
&Ne&\multicolumn{3}{c}{$0.42_{-0.04}^{+0.04}$}\\
&Mg&\multicolumn{3}{c}{$0.86_{-0.12}^{+0.14}$}\\
&Si&\multicolumn{3}{c}{$3.74_{-0.54}^{+0.60}$}\\
&Fe&\multicolumn{3}{c}{$0.94_{-0.23}^{+0.26}$}\\
\hline
\multicolumn{5}{c}{layer 1}\\
\hline
$kT_e\;(\mathrm{keV})$ &  & $0.53_{-0.06}^{+0.07} $& $0.62_{-0.04}^{+0.04}$&$0.52_{-0.04}^{+0.05}   $\\
$n_{\mathrm{e}}t\;(\mathrm{cm^{-3}\; s})$&&$3.79_{-0.67}^{+0.91} \times10^9$&$2.26_{-0.26}^{+0.32}\times10^9$&$3.41_{-0.53}^{+0.69}\times10^9$\\
norm\;(13737)\footnotemark[$\dag$]&&$9.38_{-1.26}^{+1.64} \times10^{-5}$&$1.42_{-0.14}^{+0.15}\times10^{-4}$&$1.37_{-0.15}^{+0.16}\times10^{-4}$\\
norm\;(1959)\footnotemark[$\dag$]&&$9.99_{-1.27}^{+1.55} \times10^{-5}$&$1.37_{-0.13}^{+0.14}\times10^{-4}$&$1.34_{-0.14}^{+0.15}\times10^{-4}$\\
 \hline
\multicolumn{5}{c}{layer 2}\\
\hline
$kT_e\;(\mathrm{keV})$ &  & $0.75_{-0.07}^{+0.08} $& $0.74_{-0.05}^{+0.06}$&$0.75_{-0.07}^{+0.07}   $\\
$n_{\mathrm{e}}t\;(\mathrm{cm^{-3}\; s})$&&$2.49_{-0.34}^{+0.44} \times10^9$&$2.68_{-0.29}^{+0.18}\times10^9$&$2.51_{-0.35}^{+0.43}\times10^9$\\
norm\;(13737)\footnotemark[$\dag$]&&$8.51_{-0.97}^{+1.10} \times10^{-5}$&$1.16_{-0.11}^{+0.11}\times10^{-4}$&$7.61_{-0.73}^{+0.81}\times10^{-5}$\\
norm\;(1959)\footnotemark[$\dag$]&&$8.78_{-0.96}^{+1.04} \times10^{-5}$&$1.19_{-0.10}^{+0.11}\times10^{-4}$&$7.54_{-0.72}^{+0.81}\times10^{-5}$\\
\hline
\multicolumn{5}{c}{layer 3}\\
\hline
$kT_e\;(\mathrm{keV})$ &  & $0.73_{-0.06}^{+0.07} $& $0.95_{0.07}^{+0.08}$&$0.86_{-0.08}^{+0.08}   $\\
$n_{\mathrm{e}}t\;(\mathrm{cm^{-3}\; s})$&&$2.04_{-0.34}^{+0.40} \times10^9$&$1.64_{-0.20}^{+0.25}\times10^9$&$1.94_{-0.32}^{+0.43}\times10^9$\\
norm\;(13737)\footnotemark[$\dag$]&&$8.87_{-1.25}^{+1.54} \times10^{-5}$&$8.55_{-1.00}^{+1.12}\times10^{-5}$&$5.83_{-0.68}^{+0.58}\times10^{-5}$\\
norm\;(1959)\footnotemark[$\dag$]&&$8.80_{-1.14}^{+1.44} \times10^{-5}$&$8.78_{-0.98}^{+1.10}\times10^{-5}$&$5.90_{-0.65}^{+0.81}\times10^{-5}$\\
\hline
\multicolumn{5}{c}{layer 4}\\
\hline
$kT_e\;(\mathrm{keV})$ &  & $0.82_{-0.08}^{+0.11} $& $0.95_{-0.08}^{+0.10}$&$0.89_{-0.08}^{+0.09}   $\\
$n_{\mathrm{e}}t\;(\mathrm{cm^{-3}\; s})$&&$1.91_{-0.35}^{+0.42} \times10^9$&$1.38_{-0.17}^{+0.22}\times10^9$&$1.46_{-0.32}^{+0.34}\times10^9$\\
norm\;(13737)\footnotemark[$\dag$]&&$5.30_{-0.82}^{+1.05} \times10^{-5}$&$6.78_{-0.88}^{+1.28}\times10^{-5}$&$6.20_{-0.99}^{+1.22}\times10^{-5}$\\
norm\;(1959)\footnotemark[$\dag$]&&$5.20_{-0.76}^{+0.96} \times10^{-5}$&$7.35_{-0.94}^{+1.11}\times10^{-5}$&$6.12_{-0.95}^{+1.16}\times10^{-5}$\\
\hline\hline
\multicolumn{2}{c}{c-stat/d.o.f}&\multicolumn{3}{c}{$16012.98/15564$}\\
\hline
\end{longtable}

\section{Discussion}
Our spectral analysis finds that the electron temperature increases with increasing distance from the shock front. In this section, we compare the observed electron temperature to the simplest electron heating process, the Coulomb scattering (section 4.1 -- 4.2). In addition, we note the uncertainty of ionization parameters $n_{\mathrm{e}}t$ (section 4.3), the comparison of electron density to previous studies (section 4.4) and the elemental abundance (section 4.5).

\subsection{Estimation of density and age of plasma in each region}
In the Coulomb scattering model, the ionization parameter $n_{\mathrm{e}}t$ is an important parameter as it indicates the degree of relaxation. In general, the thermal relaxation via the Coulomb scattering progresses and the electron temperature increases as $n_{\mathrm{e}}t$ increases. The value of $n_{\mathrm{e}}t$ can be estimated from the electron density and the age after the shock wave passed through, so we first estimate both of them.

We note that $n_{\mathrm{e}}t$ in our VNEI spectral models could be another parameter useful in comparison with the theoretical model, but we did not use it owing to its relatively large systematic uncertainty as described below. \citet{2013ApJ...769...64R}, for example, compared the time evolution of $n_{\mathrm{e}}t$ in VNEI model and the shock dynamical prediction from the electron density and the difference of two observation time they used, and found that two $n_{\mathrm{e}}t$ values are not consistent with each other. In addition, this spatial variation of $n_{\mathrm{e}}t$ has no correlation with the geometrical prediction from emission measure. As another example showing such uncertainty of $n_{\mathrm{e}}t$ parameter, \citet{2021ApJ...914..103S} compared SNR ages calculated from $n_{\mathrm{e}}t$ obtained in spectral modeling with those from shock dynamics and geometry and showed that they have difference about at most a factor of four. 
Here we have two $n_{\mathrm{e}} $t values, i.e., one from the spectral modeling, and the other from the emission measure and geometry.
In our case, focusing on a small region, the estimated densities and shock dynamical times are consistent with other observations as we describe later. Thus, in this work, we assume that the $n_{\mathrm{e}} t$ values calculated from the densities and shock dynamics are more suitable in comparison with the theoretical model.

The electron density is estimated from EM of our model, which is written as:

\begin{eqnarray}
%\nonumber
\mathrm{EM}=\frac{10^{-14}}{4\pi D^2}\int n_{\mathrm{e}}n_\mathrm{H}dV \;
\end{eqnarray}

\noindent where $D$ is the distance to the source, 2.18\;kpc \citep{2003ApJ...585..324W}, $n_\mathrm{H}$ is the proton density. Assuming that the electron density is equal to the proton density and their density is constant within each region, EM is proportional to the electron density squared and the volume of each region. We thus can estimate the electron density by assuming the volume of each region.

\begin{figure*}
\centering
\includegraphics[keepaspectratio,width=1.6\columnwidth]{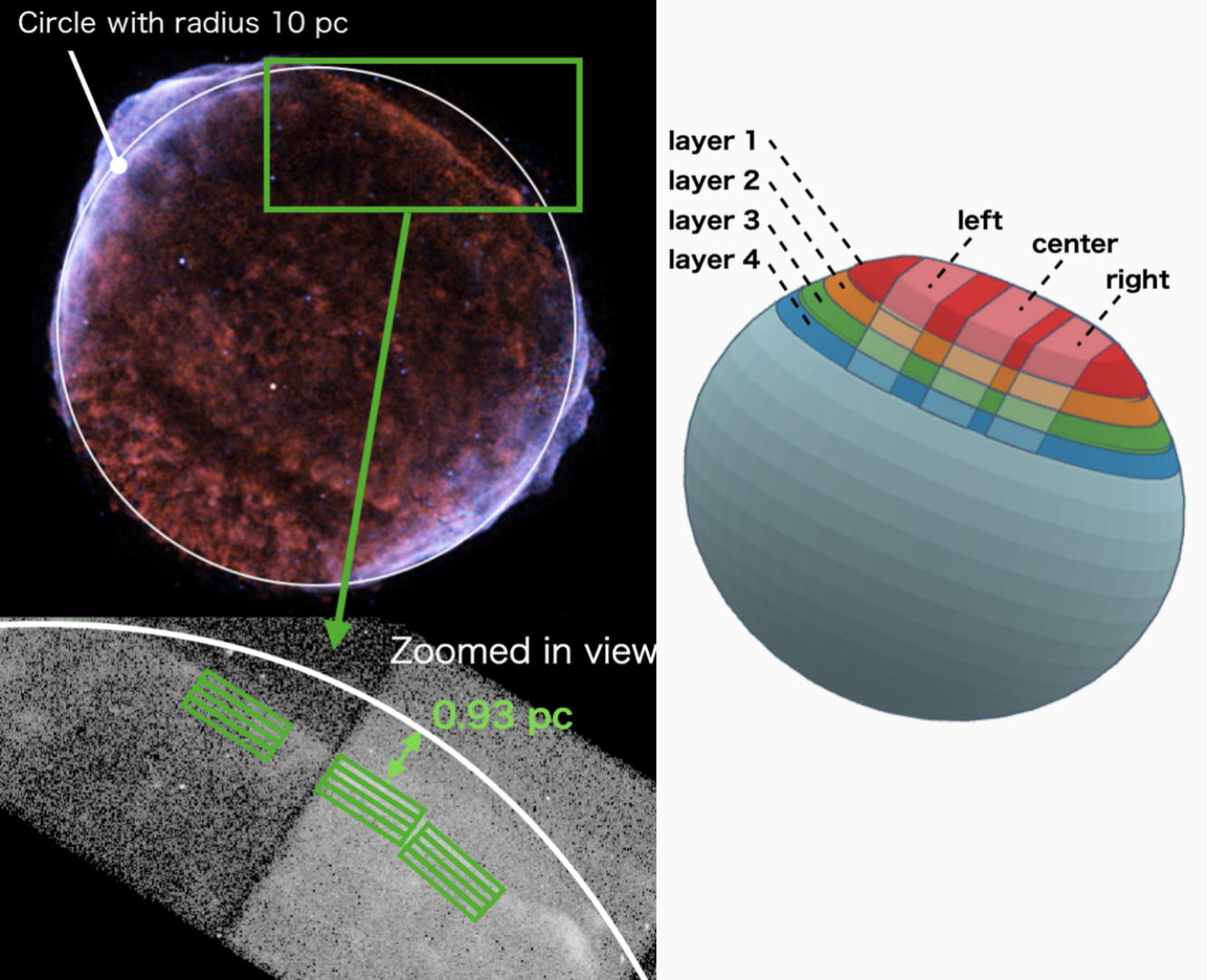}
\caption{The image of SN~1006 with a circle with a radius of 10\;pc (upper left), enlarged view of its northwestern region (lower left) and the image of the observation region (right). 12 green box regions in the lower left panel are our analysis regions (see Figure \ref{fig:region}). The shape of SN~1006 is close to a sphere with a radius of 10 pc with $\sim0.93\;\mathrm{pc}$ indentation in the northwestern region. 
Each color band in the right panel shows each observation layer (red : layer1, orange : layer2, green : layer3, blue : layer4), while translucent areas show our analysis regions.}\label{fig:sphere_lack}
\end{figure*}

To estimate the volume, we must consider the collision to H$_{\rm I}$ region in the northwestern region \citep{2022ApJ...933..157S}. Because of this collision, the shape of the northwestern region of SN~1006 has a $\sim0.93\;\mathrm{pc}$ indentation from a sphere with a radius of $10\;\mathrm{pc}$ (see the left panel of Figure \ref{fig:sphere_lack}). The volume calculated geometrically by considering this indentation like the right panel of Figure \ref{fig:sphere_lack} is shown in Table \ref{tab:estimated_density}. The electron density of each region estimated from the emission measure and the volume of each region is also shown in Table \ref{tab:estimated_density}.

\begin{longtable}[htbp]{*{4}{c}}
\caption{\mbox{The estimated volume and the mean density of each region}}
\label{tab:estimated_density}
\hline
Parameter&\multicolumn{3}{c}{Value} \\
&left&center&right\\
\endhead
\hline\hline
\multicolumn{4}{c}{layer 1}\\
\hline
volume ($\mathrm{cm}^3$)&\multicolumn{3}{c}{$5.77\times10^{57}$} \\
mean electron density (13737, $\mathrm{cm}^{-3}$)&0.299&0.357&0.361\\
mean electron density (1959, $\mathrm{cm}^{-3}$)&0.308&0.362&0.357\\
\hline
\multicolumn{4}{c}{layer 2}\\
\hline
volume ($\mathrm{cm}^3$)&\multicolumn{3}{c}{$6.18\times10^{57}$} \\
mean electron density (13737, $\mathrm{cm}^{-3}$)&0.274&0.320&0.259\\
mean electron density (1959, $\mathrm{cm}^{-3}$)&0.279&0.324&0.258\\
\hline
\multicolumn{4}{c}{layer 3}\\
\hline
volume ($\mathrm{cm}^3$)&\multicolumn{3}{c}{$6.54\times10^{57}$} \\
mean electron density (13737, $\mathrm{cm}^{-3}$)&0.273&0.268&0.221\\
mean electron density (1959, $\mathrm{cm}^{-3}$)&0.272&0.272&0.222\\
\hline
\multicolumn{4}{c}{layer 4}\\
\hline
volume ($\mathrm{cm}^3$)&\multicolumn{3}{c}{$6.94\times10^{57}$} \\
mean electron density (13737, $\mathrm{cm}^{-3}$)&0.204&0.231&0.221\\
mean electron density (1959, $\mathrm{cm}^{-3}$)&0.202&0.241&0.220\\
\hline
\end{longtable}

We estimate another important parameter, shock velocity, and its evolution. The shock speed before the collision can be estimated from the shock speed of the region except for NW. The shape of NW region has a 0.93 pc indentation from the sphere, and the shock of other region has little deviation from the sphere. This fact indicates that most of the region except for the NW region of SN~1006 does not collide with anything, and that their shock speed can be approximated by the time evolution of Sedov solution $\sim t^{-3/5}$.  
The recent shock speed in these regions is measured to be about $6000\;\mathrm{km\;s}^{-1}$ \citep{2014ApJ...781...65W}, so we estimate the time variation of shock speed without collision to H$_{\rm I}$ region as  $v_s(t)=6000\times(t/1000\;\mathrm{yr})^{-3/5}\;\mathrm{km\;s}^{-1}$, where $t$ means the age of SN~1006. In the NW region, the shock wave collided to the H$_{\rm I}$ region and suddenly decreased at the time when the collision to the H$_{\rm I}$ region started, $t'$. Before collision, the shock speed in the NW region is considered to be the same as the shock speed of the other region without colliding to the H$_{\rm I}$ region, $v_s(t)$. We also assume that the shock speed after the collision of the NW region to the H$_{\rm I}$ region is constant of recent value $v_{after}=2800\;\mathrm{km\;s}^{-1}$.

The time when the collision to the H$_{\rm I}$ region started $t^{\prime}$ can be calculated by the following equation:

\begin{eqnarray}
%\nonumber
D=\int_{1006-t^{\prime}}^{1006} (v_s(t)-v_{after}) dt \;
\end{eqnarray}

\noindent where $D$ is the indent size. Given that $D=0.93\;\mathrm{pc}$ for the northwestern region of SN~1006, the time of the collision $t'$ is calculated to be about 250 yr, or $0.18\;\mathrm{pc}$ from the shock front. This estimation indicate that the plasma in the layer 1 was heated by the forward shock after it collided with the H$_{\rm I}$ region, and that plasmas in layers 2, 3, and 4 were heated by the forward shock before the collision with the H$_{\rm I}$ region.

\begin{figure*}
\centering
\includegraphics[keepaspectratio,width=1.6\columnwidth]{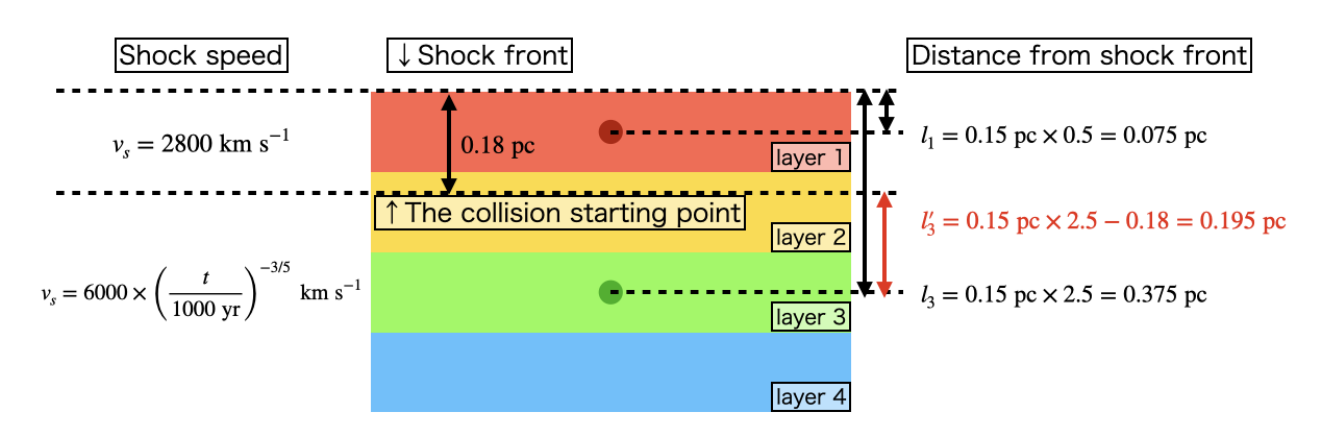}
\caption{Assumption for calculating the plasma age of each layer. The shock speed means the speed at the time the shock wave passed through (see fifth paragraph of section 4.1). The collision starting point is calculated from the indentation of the northwestern region of SN~1006 (see Figure \ref{fig:sphere_lack}). The distance of each layer from the shock is defined to the center of each layer.}\label{fig:shock_speed_figure}
\end{figure*}

The plasma age of each layer is estimated from the distance from the shock front divided by the shock speed assumed above. We show a schematic summarizing factors for calculating the plasma age in Figure \ref{fig:shock_speed_figure}.
We note that the plasma we observed is the ISM plasma swept up by the shock wave. These plasmas move in the same direction of the shock wave with the speed 3/4 of the shock speed \citep{1994ApJ...420..721H}, so the plasma age of each layer $l_i$ ($i=1, 2, 3$ and $4$) is four times larger than the value obtained by dividing the distance from the shock front by the shock speed. Specifically, the age of the center of layer 1, where the shock speed is constant at $2800\;\mathrm{km\;s}^{-1}$, is calculated as $(0.15\;\mathrm{pc})/2/2800\;\mathrm{km\;s}^{-1}\;\times4\;\sim\;112\;\mathrm{yr}$. 
For calculating the age of layers 2, 3 and 4, we have to consider the sudden change of the shock speed due to the collision and the time variation of the shock speed before collision. For 250 years after the collision to the H$_{\rm I}$ region, the shock velocity is assumed as constant, and pass through $0.18\;\mathrm{pc}$. For the distance from the shock front over $0.18\;\mathrm{pc}$, the shock wave passed through with the shock velocity before the collision $v_s(t)=6000\times(t/1000\;\mathrm{yr})^{-3/5}\;\mathrm{km\;s}^{-1}$, which $t$ means the age of SN~1006. Defining the age before the collision to the H$_{\rm I}$ region, the distance which the shock passed through before the collision $l_i^{\prime}$ is calculated as below:

\begin{eqnarray}
%\nonumber
l_i^{\prime}\;(\mathrm{pc})=\int^{1000-250\;(\mathrm{yr})}_{t_i}v_s(t)/4\;dt=0.24\times(750^{0.4}-t_i^{0.4})\;(\mathrm{pc})\;
\end{eqnarray}

\noindent For example, in the case of layer 3, the distance which the shock passed through before the collision $l_3^{\prime}$ is calculated as $l_3=0.16\times2.5-0.18=0.22\;\mathrm{pc}$. By using the equation above, we can calculate $t_3=109\;\mathrm{yr}$, so the plasma age of layer 3 is estimated to be $359\;(=109+250)\;\mathrm{yr}$. The results of the plasma age of all layers are shown in Table \ref{tab:estimated_age}.

\begin{table}
\tbl{Plasma age of observation region and estimated shock speed}{%
\begin{tabular}{ccc}
\hline
Region&plasma age (yr)\footnotemark[a]&Shock velocity ($\mathrm{km\;s}^{-1}$)\footnotemark[b] \\
\hline\hline
layer 1&112 & $2.8\times10^3$\\
layer 2&269 & $7.2\times10^3$\\
layer 3&359 & $7.8\times10^3$\\
layer 4&445 & $8.5\times10^3$\\
\hline
\end{tabular}}\label{tab:estimated_age}
\begin{tabnote}
 \footnotemark[a] Plasma age of each layer is represented by the value of center of each region. \\
 \footnotemark[b] The shock velocity before collision can be approximated by the time evolution of Sedov solution $\sim t^{-3/5}$. The shock velocity after the collision as constant of recent value $2800\;\mathrm{km\;s}^{-1}$. 
%\item[$b$] 
\end{tabnote}
\end{table}

To calculate the electron temperature relaxation derived from the Coulomb scattering model, we need to estimate the immediate postshock electron temperature. The simplest estimation is the Rankine-Hugoniot relation $kT_e=\frac{3}{16}m_iv_s^2$, where $m_i$ is the mass of each particle and $v_s$ is the shock speed. In this paper, we consider the case of $v_s=2800\;\mathrm{km\;s}^{-1}$ (after the collision) and $v_s=6000\;\mathrm{km\;s}^{-1}$ (lower limit of the shock speed before the collision).

\begin{figure}
\centering
\includegraphics[keepaspectratio,width=1.0\columnwidth]{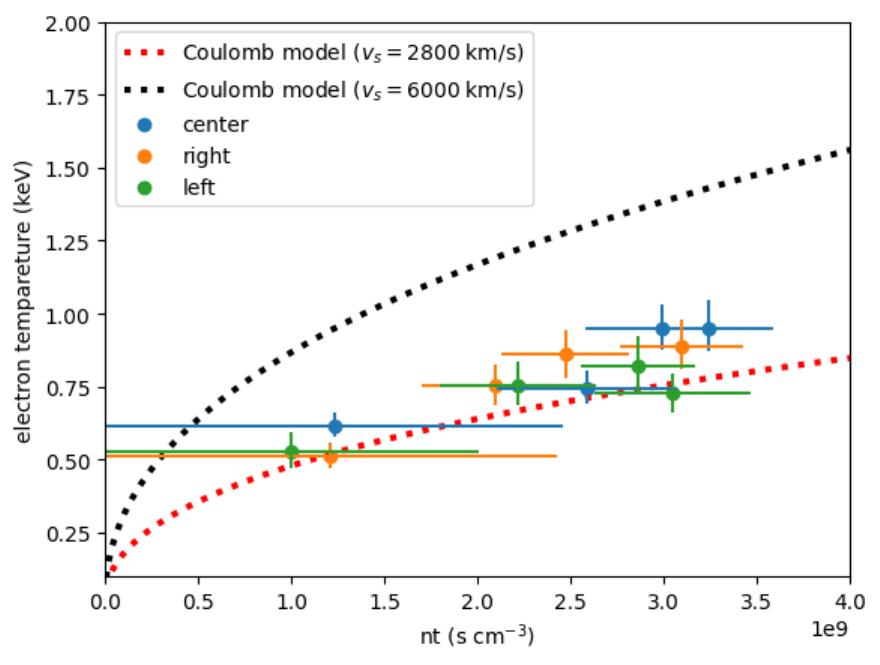}
\caption{Comparison the observed electron temperature to that estimated from the Coulomb scattering model heating. The shock speed is assumed to be $2800\;\mathrm{km/s}$ for the red line and $6000\;\mathrm{km/s}$ for the black line. Note that the parameter $n_{\mathrm{e}}t$ is calculated from emission measure and shock dynamics, not from spectral modeling. The initial temperature ratio $\beta$ is assumed to be the minimum value, $\beta=m_e/m_p$.}\label{fig:comp_coulomb}
\end{figure}

The result of the comparison between the observed electron temperature and the Coulomb scattering model is shown in Figure \ref{fig:comp_coulomb}.
In this comparison, the parameter $n_{\mathrm{e}}t$ of the data is calculated as the product of the density $n_\mathrm{e}$ and elapsed time after the shock passage $t$. We evaluate the error range of $n_{\mathrm{e}}t$ considering the possible range of the elapsed time $t$. We assume that distances from the shock front to the farthest and nearest edges of each region yield the maximum and minimum values of the elapsed time $t$, respectively. 
In the layer 1, where the shock wave passes through after the collision to the H$_{\rm I}$ region with a shock speed of $2800\;\mathrm{km\;s}^{-1}$, the observation and the Coulomb scattering model are consistent with each other. On the other hand, the electron temperatures in layers 2, 3 and 4, where the shock wave passes through before the collision and the shock speed is assumed to be more than $6000\;\mathrm{km\;s}^{-1}$, are significantly lower than that estimated from the Coulomb scattering model.

\subsection{Possible origins of the lower temperatures compared with the Coulomb scattering model}
In this section, we discuss the discrepancy between our observed electron temperatures in some observation region and the Coulomb scattering model.
 
The first possibility is a bias due to the mixture of high-temperature and low-temperature plasmas. The outer regions should have lower electron temperatures than the inner regions because the outer regions seem to have been heated by slower shocks and the electron heating does not proceed as in the inner regions.  Such a temperature gradient would make the apparent temperature in the inner regions lower than the actual values due to projection effects. One of the ways to consider this effect is to make a fitting model that reflects these detailed density and temperature structures, although lack of the statistics of the current dataset prevents us from resolving this effect.

Alternatively, the real temperature could be reduced in the inner regions. If the high-temperature ISM in the inner region and relatively low-temperature ISM in the outer region are mixed in turbulence, with each other to some extent, we expect a decrease in the ISM electron temperature in the inner region from that of the Coulomb scattering model. 

To examine this explanation, we need to understand the degree of ionization and turbulence, which we can estimate from the width of characteristic X-ray lines. They would tell us about the bulk and thermal motion velocities. Better understanding of the plasma turbulence will be realized with calorimeter missions such as XRISM \citep{2020SPIE11444E..22T} and Athena \citep{2013arXiv1306.2307N}.

Another possibility is the energy leakage to particle acceleration. The SNR shock wave is one of the environments where the cosmic rays are actively accelerated (e.g. \cite{2005ApJ...621..793B}). Such an efficient acceleration can steal energy from the shocked plasma, which makes the downstream temperature lower than that in the case without energy loss. The comparison of the temperature change between the northwestern and the northeastern limbs, where efficient acceleration is reported, will resolve this scenario.

Note that our result, lower temperatures compared with the Coulomb scattering model, is different from previous results by \citet{2014ApJ...780..136Y}, for example. Although it is unclear what makes such a different result, we stress that the situation of our region and the  region analyzed be \citet{2014ApJ...780..136Y} is different: our results are temperature changes of ISM in high density gradient, whereas the results by \citet{2014ApJ...780..136Y} are for ejecta without high density gradient.

\subsection{On the uncertainty of $n_{\mathrm{e}}t$}
\begin{figure}
\centering
\includegraphics[keepaspectratio,width=1.0\columnwidth]{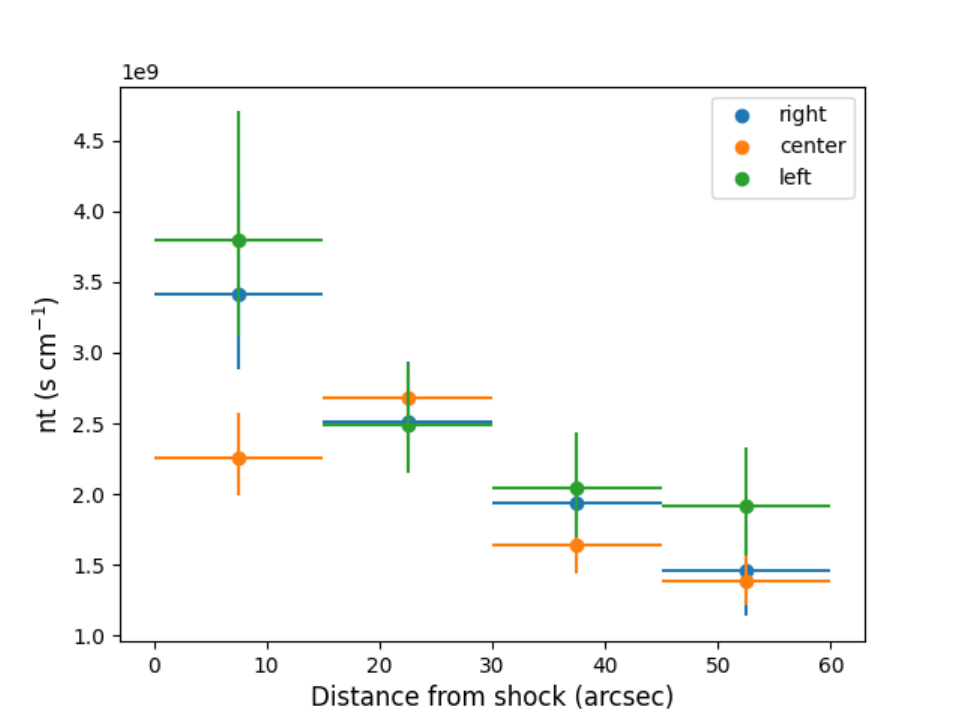}
\caption{Ionization parameter $n_{\mathrm{e}}t$ from the best-fit VNEI models as a function of the distance from the shockfront}\label{fig:nt_model}
\end{figure}

\begin{figure}
\centering
\includegraphics[keepaspectratio,width=1.0\columnwidth]{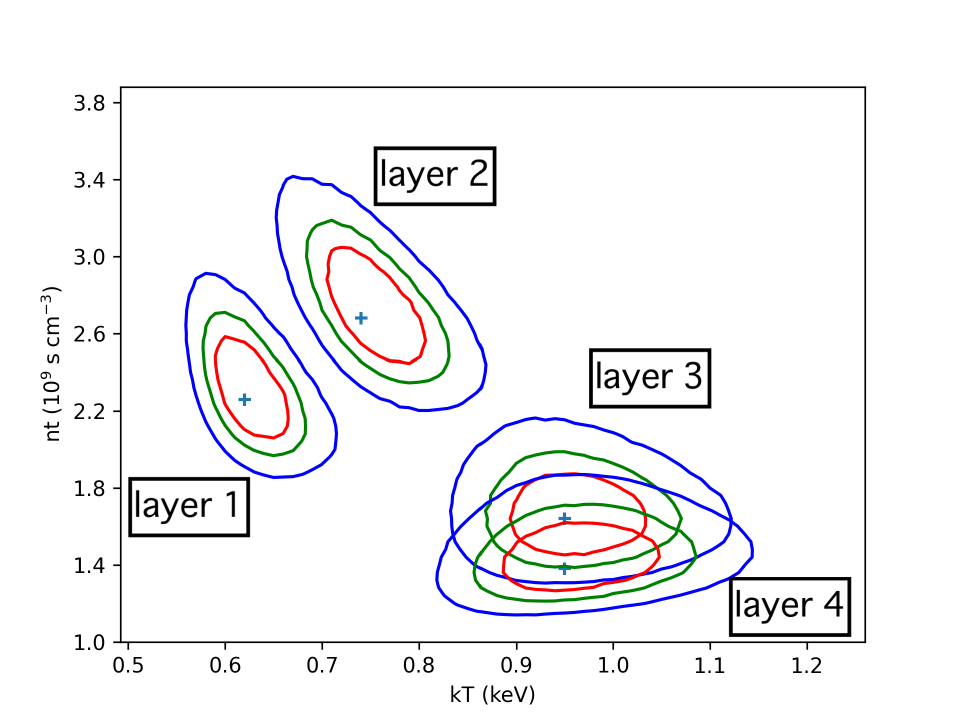}
\caption{The $kT$--$n_{\mathrm{e}}t$ contours of the center region. Each red, green and blue line shows the 68\%, 90\% and 99\% error range, respectively. Note that the parameter $n_{\mathrm{e}}t$ is from the best-fit VNEI models.}\label{fig:kT_nt_contour}
\end{figure}

While we estimate $n_{\mathrm{e}}t$ from the emission measure and the time after the shock heating, $n_{\mathrm{e}}t$ is also derived from the spectral fitting with the VNEI model. However, as mentioned in the beginning of the section 4.1, 
$n_{\mathrm{e}}t$ in the VNEI model decreases toward inside, although the opposite trend is expected theoretically as shown in Figure \ref{fig:nt_model}. 
 The rapid decrease of the electron density in observation region can cause this trend, but this is not consistent with what is expected from the spatial variation of the emission measure. The coupling between $kT$ and $n_{\mathrm{e}}t$ in the spectral analysis can cause this suspicious behavior, but as shown in Figure \ref{fig:kT_nt_contour}, $kT$ and $n_{\mathrm{e}}t$ from the spectral fitting are not strongly coupled as shown in Figure \ref{fig:kT_nt_contour}. We also fix all of $n_{\mathrm{e}}t$ in the spectral fitting to be the value estimated in section 4.1 and fit again, but significant residuals appear especially in thermal energy range (0.5 -- 2.0 $\mathrm{keV}$). Note that we cannot simply compare Figure \ref{fig:kT_nt_contour} to the Coulomb scattering model shown in Figure \ref{fig:comp_coulomb} because of this uncertainty in $n_{\mathrm{e}}t$, even if they appear to be consistent with each other. In fact, the contours between $kT$ and $n_{\mathrm{e}}t$ in both right and left region are significantly different from the Coulomb scattering model in Figure \ref{fig:comp_coulomb}.

It is not easy to figure out the cause of the mysterious behavior of the ionization parameter. However, one possibility is the transport of inner plasma outward by turbulence. This transport could bring up highly ionized plasmas deep inside post shock regions to regions right behind the shock, so that the ionization parameter we observed near the shock becomes larger apparently. This plasma mixing could also cause the averaging of the electron temperatures as well as the ionization parameter, which can explain the slower-than-expected temperature evolution as shown in Figure \ref{fig:kT}.

Finally, we have made the consistency check of the assumption that $n_{\mathrm{e}}t$ is linked between the two observation in 2001 and 2012. $n_{\mathrm{e}}t$ in two observations is considered to change by multiplying the electron density and the time between these observations. This value is $4.5\times10^7\;-\; 3.9\times10^8\;\mathrm{cm}^{-3}\;\mathrm{s}$. These values are less than 10 \% of the $n_{\mathrm{e}}t$ in our analysis. This result justifies our assumption in analysis of $n_{\mathrm{e}}t$ constant in both data observed in 2001 and 2012.

\subsection{Interpretation of the pre-shock density}
It is important to compare our measured electron density to the pre-shock electron density estimated in previous studies, $n_{\mathrm{e}}=0.03\;-\;0.40\;\mathrm{cm}^{-3}$ (\cite{2007A&A...475..883A}, \cite{2016MNRAS.462..158L}, \cite{2007ApJ...659.1257R}, \cite{2003ApJ...586.1162L}). The pre-shock density in our analysis is estimated from the electron density of layer 1. Layer 1 is mostly occupied by the high density region just after the H$_\mathrm{I}$ region, so the pre-shock density can be simply estimated by dividing the electron density in layer 1 by 4. The value of the estimated pre-shock density is $0.074-0.091\;\mathrm{cm}^{-3}$, which is consistent with the density of the previous studies listed above.

On the other hand, these pre-shock densities are not consistent with the proton density of the H$_\mathrm{I}$ region, larger than $12\;\mathrm{cm}^{-3}$ \citep{2022ApJ...933..157S}. We note that this density is considered as a lower limit, as it is estimated based on the assumption that the mass of HI shell of about 1000 M$_\odot$ is uniformly distributed over the present volume of the remnant, but in reality the same mass is accumulated within a thin shell (see \cite{2022ApJ...933..157S} for more detailed information).
Layer 1 is, at least apparently, just after the shock, so it is reasonable to assume that the shock is currently passing through the H$_\mathrm{I}$ region, so this discrepancy in density is puzzling. \citet{2022ApJ...933..157S} claims that the X-ray emitting region still lies in the cavity. Under this argument, the pre-shock density estimated in the previous X-ray studies is interpreted as the cavity density. However, this interpretation also contradicts a simple hydrodynamic shock-cloud interaction model \citep{1994ApJ...420..721H}. In the model of \citet{1994ApJ...420..721H}, decrease in shock velocity depends on the density ratio between the H$_\mathrm{I}$ region and the cavity. In the case of the northwestern region of SN~1006, the shock velocity decreases from $6000\times(750/1000)^{-3/5}\;\sim7130\;\mathrm{km\;s}^{-1}$ to 2800 $\mathrm{km\;s}^{-1}$, so the density ratio between the H$_\mathrm{I}$ region and the cavity is estimated to be $\sim 21.3$, leading to a cavity density of $0.56\;\mathrm{cm}^{-3}$. This density is much larger than that we measured ($0.074-0.091\;\mathrm{cm}^{-3}$) as well as past observations ($0.03-0.4\;\mathrm{cm}^{-3}$).

Although the cause of this discrepancy in pre-shock density is not understood at this time, there are a number of  factors that cause the uncertainty in the density that may lead to this discrepancy. The most significant factor is the assumption of the volume of each region. Since we cannot measure the actual volume, we cannot rule out the possibility that the volume assumed here may be grossly incorrect. If the actual volume of the observed region is about 100 times smaller, the electron density is estimated to be 3.1 times larger and the pre-shock density discrepancy is greatly improved. The uncertainty in the distance of SN~1006 from the Earth also affects the uncertainty in the volume assumption. We assume the distance of SN~1006 to be 2.18\;kpc \citep{2003ApJ...585..324W}, but another study estimated it to be 1.5\;kpc \citep{2017hsn..book...63K}. If the actual distance to SN~1006 is 1.5\;kpc, the volume we assumed in section 4.1 is about 70\% smaller, leading to a density 1.8 times higher. The 3-D density structure also affects the uncertainty of the pre-shock density estimation. In our estimation, we assume that the electron density in each region is constant. However, the actual density structure of each region consists mainly of three parts: the high density region just after H$_\mathrm{I}$ region, the medium density region where the reflected wave has passed, and the low density region where the reflected wave has not yet reached. The observed spectra are superpositions of these, so if the density and other parameters of each region are to be determined in more detail, simultaneous modeling is required. With the current statistics, however, we find it difficult to obtain reasonable results.

\subsection{Comments on possible spatial variation of metal abundances}
In our analysis, the metal abundances are fixed to constant values in all the analysis regions. The result of fitting reproduces the observed spectra well, but slight residuals at the Ne and Mg lines remain in some layers, which may imply that the Ne and Mg abundances change within the analysis regions. We try a spectral fit with the Ne and Mg abundances free, but these abundances show no statistically significant changes. We expect that future satellites such as XRISM \citep{2020SPIE11444E..22T} and Athena \citep{2013arXiv1306.2307N} will enable the resolution of spatial variations in metal abundances, with their excellent energy resolutions.

\section{Conclusion}
We analyzed the northwestern region of SN~1006 with Chandra and measured a spatial variation of the electron temperature just behind the shock for the first time. We selected three regions and then divided them into four layers along the shock normal, with a thickness of $15^{\prime\prime}$ or 0.16 pc each. All the spectra were well explained with a single ionizing plasma model. 
The electron temperature increases toward downstream from 0.56 to 0.89 keV on a length scale of 60 arcsec. Several possibilities are suggested to make this tendency, such as thermal relaxation processes and plasma mixture due to turbulence, projection effects, etc. We expect  a breakthrough in separating these potential contributions with future satellites with better statistic and higher energy resolutions such as XRISM and Athena.

\section{Acknowledgement}
This work is partly supported by JSPS Grants-in-Aid for Scientific Research 19K03908 (AB), 23H01211 (AB), 20H00174, 21H01121 (SK)  and 21J00031 (HS).

\end{document}